\documentclass[10pt]{iopart}
\usepackage{graphicx}
\eqnobysec

\newcommand{\beq}{\begin{equation}}
\newcommand{\beqa}{\begin{eqnarray}}
\newcommand{\eeq}{\end{equation}}
\newcommand{\eeqa}{\end{eqnarray}}

\newcommand{\abs}[1]{\vert#1\vert}
\newcommand{\B}{c}
\newcommand{\bigmean}[1]{\left\langle#1\right\rangle}
\newcommand{\C}{{\cal C}}
\newcommand{\comport}[2]{\mathrel{\mathop{#1}\limits_{#2}^{}}}
\newcommand{\de}{^{(I\!I)}}
\newcommand{\dpar}{\partial}
\newcommand{\dq}{\!{\frac{\d^D\q}{(2\pi)^D}}}
\newcommand{\E}{{\bf e}}
\newcommand{\euler}{{\bf C}}
\newcommand{\erf}{\mathop{\rm erf}\nolimits}
\newcommand{\erfc}{\mathop{\rm erfc}\nolimits}
\newcommand{\eps}{\varepsilon}
\newcommand{\fd}{fluc\-tu\-a\-tion-dis\-si\-pa\-tion }
\newcommand{\frad}[2]{\displaystyle{\displaystyle#1\over\displaystyle#2}}
\renewcommand{\H}{{\cal H}}
\newcommand{\K}{{\bf K}}
\newcommand{\mean}[1]{\langle#1\rangle}
\newcommand{\q}{{\bf q}}
\renewcommand{\Re}{\mathop{\rm Re}\nolimits}
\renewcommand{\SS}{{\cal S}}
\newcommand{\un}{^{(I)}}
\newcommand{\V}{{{\bf V}}}
\newcommand{\w}{\widehat}
\newcommand{\X}{{\cal X}}
\newcommand{\x}{{\bf x}}
\newcommand{\y}{{\bf y}}
\newcommand{\0}{{\bf 0}}
\newcommand{\m}{{\hat\mu}}
\newcommand{\n}{{\hat\nu}}

\newcommand{\aalpha}{{\rm alpha}}
\newcommand{\as}{{\rm as}}
\newcommand{\asym}{{\rm asym}}
\newcommand{\bbeta}{{\rm beta}}
\renewcommand{\d}{{\rm d}}
\renewcommand{\e}{{\rm e}}
\newcommand{\eq}{{\rm eq}}
\newcommand{\F}{{\rm F}}
\newcommand{\ii}{{\rm i}}
\renewcommand{\L}{{\rm L}}
\newcommand{\reg}{{\rm reg}}
\newcommand{\sg}{{\rm sg}}
\newcommand{\st}{{\rm stat}}

\begin{document}

\title{Asymmetric Langevin dynamics for the ferromagnetic spherical model}

\author{C Godr\`eche and J M Luck}

\address{Institut de Physique Th\'eorique, CEA Saclay and CNRS,
91191~Gif-sur-Yvette cedex, France}

\begin{abstract}
The present work pursues the investigation of the role
of spatial asymmetry and irreversibility on the dynamical properties of spin systems.
We consider the ferromagnetic spherical model
with asymmetric linear Langevin dynamics.
Such an asymmetric dynamics is irreversible, i.e., breaks detailed balance,
because the principle of action and reaction is violated.
The fluctuation-dissipation theorem therefore no longer holds.
The stationary state is however still Gibbsian,
i.e., the weights of configurations are given by the Boltzmann factor
corresponding to the ferromagnetic Hamiltonian.
The model is exactly solvable in any dimension,
enabling an analytical evaluation of time-dependent observables.
We show the existence of two regimes of violation
of the fluctuation-dissipation theorem in the nonequilibrium stationary state:
a regime of weak violation where the stationary fluctuation-dissipation ratio is finite
but less than unity and varies continuously with the asymmetry,
and a regime of strong violation where the fluctuation-dissipation ratio
vanishes asymptotically.
This phenomenon was first uncovered
in the asymmetric kinetic Ising chain.
The present results suggest that this novel kind of dynamical transition in nonequilibrium stationary states might be quite general.
We also perform a systematic analysis of several regimes of interest,
either stationary or transient,
in various dimensions and in the different phases of the model.
\end{abstract}

\eads{\mailto{claude.godreche@cea.fr},\mailto{jean-marc.luck@cea.fr}}

\maketitle

\section{Introduction}
\label{intro}

In the past decades many studies have been devoted to the dynamics
of ferromagnetic spin systems.
If such a system is quenched from a disordered initial state,
it relaxes, at high temperature, towards its paramagnetic equilibrium state,
whereas, in the low-temperature ordered phase, equilibrium is never reached
and the system undergoes a non-stationary coarsening process~\cite{bray}.

Both situations are illustrated by the exactly solvable
Glauber-Ising chain~\cite{glau}.
At finite temperature, where the relaxation time is finite,
the convergence to the paramagnetic equilibrium state can be monitored by the exact determination of the two-spin correlation functions,
either for equal or unequal times.
When the system has reached equilibrium, its response to a small perturbation
is simply related to the two-time correlation function, which
is the content of the \fd theorem~\cite{glau}.
In the low-temperature regime, where the relaxation time is diverging,
the system undergoes coarsening, with scaling and universality properties
which have been characterized in more recent studies~\cite{bray,gl1D,zan,sollich,malte}.
In higher dimensions, since no exact result prevails, the present understanding of the properties of kinetic Ising models has been achieved by
numerical simulations, scaling arguments
and field-theoretical methods~\cite{bray}.

For kinetic Ising models with non conserved (Glauber) dynamics, a line of investigation which has been almost completely overlooked until recently is the role of spatial asymmetry in the dynamics.
This is in contrast with the case of Ising models with conserved (Kawasaki) dynamics
in the presence of a bias due to an external field,
which have been thoroughly investigated, mostly numerically~\cite{kls}.
In the context of Glauber dynamics,
asymmetry means that the flipping spin is not equally influenced by all its neighbours.
Such an asymmetric dynamics is therefore irreversible,
because the principle of action and reaction is violated
and detailed balance no longer holds.
In a recent past, two studies have been devoted to
Ising models with non conserved asymmetric dynamics~\cite{kun,stau}.
Reference~\cite{kun} gives examples of rate functions for the kinetic Ising model in
one and two dimensions with totally asymmetric dynamics leading to a Gibbsian
stationary state
(in the sense that the weights of configurations are given by the Boltzmann factor
corresponding to the ferromagnetic Hamiltonian).
Reference~\cite{stau} investigates, by numerical means, the possible existence of a
phase transition for the kinetic Ising model, in dimensions two to five,
when the usual heat-bath rate is modified by truncating the local field,
keeping a subset of influential spins only.
The dynamics thus obtained neither fulfills detailed balance,
nor global balance (with respect to the ferromagnetic Hamiltonian),
and therefore leads to an unknown stationary state, except in low dimension,
where the problem can be easily analyzed~\cite{gb2009}.

The works~\cite{kun,stau} motivated the systematic study presented in~\cite{gb2009}.
In particular, the examples given by K\"{u}nsch~\cite{kun}
raised the following questions:
{\it To what extent are these examples unique?
Can they be extended to lattices in dimension higher than two?}
As demonstrated in~\cite{gb2009},
as long as the dynamics is not totally asymmetric,
the space of parameters defining the rate function
of irreversible Gibbsian Ising models in one and two dimensions is large.
However, imposing total asymmetry of the dynamics yields a unique solution
(up to a global time scale), for all the examples considered
(linear chain, square and triangular lattices).
The answer to the second question is presumably negative: there is no such Gibbsian irreversible dynamics for the cubic lattice and one can argue that there are neither totally asymmetric Gibbsian dynamics for lattices of coordination $z\ge8$~\cite{gb2009}.

Such Gibbsian asymmetric kinetic models,
where the flipping spin is not equally influenced by all its neighbours,
provide benchmarks to study the phys\-ical consequences of irreversibility, in particular the properties of the resulting non\-equi\-li\-brium stationary state.
For instance, for the Ising chain, a one-parameter family of asymmetric dynamics
can been shown to keep the exact solvability of the symmetric Glauber dynamics,
allowing a detailed analysis of its dynamical features~\cite{cg2011}.
One of the most striking results is the existence of two regimes
of violation of the \fd theorem in the nonequilibrium stationary state:
when the asymmetry parameter is less than a threshold value,
there is weak violation of the \fd theorem, with a finite asymptotic \fd ratio,
whereas this ratio vanishes above the threshold~\cite{cg2011}.
At low temperature, the asymptotic \fd ratio depends on the strength of the asymmetry~\cite{gl1D}, and in particular vanishes at long times, as soon as the asymmetry is present, while in the absence of asymmetry it takes a universal value equal to $1/2$~\cite{gl1D}.
For the two-dimensional Ising model
with the asymmetric dynamics introduced in~\cite{gb2009}
numerical simulations show many features in common with the one-dimensional case, in particular the existence of two regimes of violation of the \fd theorem~\cite{gp}.
Irreversibility also implies a non-vanishing entropy production rate
in the stationary state, which can be exactly computed for irreversible Gibbsian Ising models since the stationary measure is known~\cite{maes,oliv}.

In the present work we pursue the investigation
of the role of spatial asymmetry and irreversibility on the dynamics of spin systems,
by considering the ferromagnetic spherical model with asymmetric Langevin dynamics.
We follow the thread of our earlier investigations
of the spherical model with symmetric Langevin dynamics~\cite{gl2D}
and of the asymmetric kinetic Ising chain~\cite{cg2011}.
The outcomes of the present study turn out to share
many common features with those of the latter reference.
In particular we show that the phenomenon described above of two regimes of violation of the \fd theorem in the stationary state also exists for the spherical model, in any dimension, and thereby acquires a status of greater generality.

The spherical model has been introduced long ago by Berlin and Kac~\cite{berlin},
as an attempt to simplify the Ising model.
It is also known to be equivalent to the $n\to\infty$ limit of the $\mbox{O}(n)$
Heisenberg model~\cite{stanley}.
Its thermodynamic properties can be studied exactly in any dimension.
The model however possesses non-trivial critical properties~\cite{berlin,baxter}.
The spherical model has another significant advantage in the present context:
its time-dependent properties under symmetric Langevin dynamics can be investigated
by analytical means, in all the phases of the model
and in any dimension~\cite{ronca,janssen,cug,horner,gl2D}.
Other aspects of the dynamics of the spherical model have been explored, such as
the presence of a constant magnetic field,
long-range or competing interactions,
finite-size and surface effects, a temperature-dependent bath,
and a correlated initial state~\cite{various}.
The dynamics of spherical spin-glass models has also received much attention,
especially in the mean-field geometry~\cite{mfsg}.

The spherical model with asymmetric dynamics has also been considered in previous studies.
One should first mention the study
of the dynamics of a mean-field spherical spin glass with random asymmetric bonds~\cite{cs},
inspired by neuronal networks with asymmetric synaptic couplings.
More recently, Langevin dynamics for the ferromagnetic spherical model
with an asymmetry has been considered in~\cite{entropy}.
The latter work was however limited to an evaluation of the entropy production rate
in the stationary state.
Instead, the main focus of the present work is on the dynamical properties of the model, at stationarity or out of stationarity, in particular through an investigation of the two-time sector
of the dynamics, and especially of the fluctuation-dissipation ratio.

The setup of the paper is as follows.
The spatially asymmetric linear Langevin dynamics
is introduced in section~\ref{gene},
and shown to be characterized by a velocity vector~$\V$.
In section~\ref{static} we show why the nonequilibrium stationary state
reached by this dynamics is Gibbsian,
in the sense that the weights of configurations are the Boltzmann weights
corresponding to the ferromagnetic Hamiltonian.
We also give a reminder of some features and properties necessary for the sequel:
critical temperature, relaxation time, equal-time correlations.
In section~\ref{two} we derive general expressions for dynamical quantities
in the two-time sector.
As the dynamics is irreversible for any non-zero $\V$,
the \fd theorem relating the two-time correlation and response functions
no longer holds, even in the stationary state.
We derive a few properties of the nonequilibrium stationary state in section~\ref{stat},
including the dispersion of the \fd ratio in Fourier space and the entropy production rate.
We then show that there exist two
distinct regimes of violation of the \fd theorem:
a regime of weak violation at small $\V$,
where the stationary \fd ratio is finite but less than unity,
and a regime of strong violation at large $\V$,
where the \fd ratio falls off exponentially with time.
This is demonstrated in one dimension~(section~\ref{stat1})
and in higher dimension~(section~\ref{stathigh}).
The next three sections contain a detailed investigation of the transient properties
of the model, before the nonequilibrium stationary state sets in.
We successively consider the low-temperature scaling regime of the one-dimensional model
(section~\ref{t1}) and the vicinity of the critical point in higher dimension,
where a continuum description makes the analysis simpler,
both in the classical regime ($D>4$) in section~\ref{tcl}
and in the non-classical one ($2<D<4$) in section~\ref{tfl}.
In all these situations,
the effect of an asymmetry on the dynamical critical properties of the model
bears strong similarities with the case of the asymmetric kinetic Ising chain.
The \fd ratio has a scaling form in two variables,
respectively related to the temperature difference $T-T_c$
and to the asymmetry parameter $\V$.
Finally, the asymmetric dynamics of the low-temperature ferromagnetic phase
is investigated in section~\ref{ferro};
it exhibits two successive regimes (beta and alpha),
in agreement with the phenomenology of glassy physics.
The quantitative behavior of the two-time correlation function
and of the corresponding \fd ratio is shown to be affected by the irreversibility of the dynamics.

\section{The model}
\label{gene}

The definition of the ferromagnetic spherical model is well known.
Consider a lattice in arbitrary dimension~$D$,
chosen to be hypercubic for simplicity.
The spins $S_\x$, sitting at the vertices~$\x$,
are real variables obeying the spherical constraint
\beq
\sum_\x S_\x^2=N,
\label{constr}
\eeq
where~$N$ is the number of spins in the system,
which is supposed to be finite for the time being.
The Hamiltonian of the model reads
\beq
\H=-\sum_{(\x,\y)}S_\x S_\y,
\label{ham}
\eeq
where the sum runs over pairs of neighboring sites.

We first remind the rules of the symmetric Langevin dynamics, before addressing the case of asymmetric dynamics, which is the focus of the paper.

\subsection{Reversible (symmetric) Langevin dynamics}

Hereafter we assume that the system is homogeneous,
i.e., invariant under spatial translations.
This holds for a finite sample with periodic boundary conditions,
and for the infinite system.
Furthermore, along the lines of earlier studies of dynamical properties,
we shall consider the {\it mean spherical model},
where the constraint~(\ref{constr}) is imposed on average,
becoming thus the mean spherical constraint~\cite{lw}:
\beq
\mean{S_\x(t)^2}=1.
\label{cave}
\eeq
Throughout the following, we consistently use the notations of~\cite{gl2D}.

Symmetric Langevin dynamics is defined by the stochastic differential equation
\beq
\frac{\d S_\x}{\d t}=F_\x-\lambda(t)S_\x+\eta_\x(t),
\label{langen}
\eeq
where $\eta_\x(t)$ is a Gaussian white noise with covariance
\beq
\mean{\eta_\x(t)\eta_\y(t')}=2T\,\delta_{\x,\y}\,\delta(t-t'),
\eeq
and $\lambda(t)$ is a Lagrange multiplier ensuring
the constraint~(\ref{cave}) at any time~$t$.
We choose for convenience to parametrize $\lambda(t)$ as
\beq
\lambda(t)=2D+z(t).
\label{zdef}
\eeq
Finally, $F_\x$ is the force exerted on $S_\x$ by the other spins, i.e.,
\beq
F_\x=-\frac{\dpar\H}{\dpar S_\x}=\sum_{a=1}^D(S_{\x-\E_a}+S_{\x+\E_a}),
\label{fgrad}
\eeq
where the $2D$ neighbors of $\x$
are $\y=\x\pm\E_a$, and $\E_a$ is a unit vector in direction $a=1,\dots,D$.
The Langevin equation~(\ref{langen}) thus reads
\beq
\frac{\d S_\x}{\d t}
=\sum_{a=1}^D(S_{\x-\E_a}+S_{\x+\E_a}-2S_\x)-z(t)S_\x+\eta_\x(t).
\label{lansym}
\eeq
This dynamics is reversible and obeys detailed balance with respect to
the Hamil\-ton\-ian~(\ref{ham}) at any temperature $T$; it drives the spherical model to the corresponding equilibrium state,
at least in the paramagnetic phase ($T>T_c$).

\subsection{Irreversible (asymmetric) Langevin dynamics}

We define the most general asymmetric Langevin dynamics by dissymetrizing the roles
of the `left' neighbors ($\y=\x-\E_a$) and of the `right' neighbors ($\y=\x+\E_a$),
while keeping the linearity of the Langevin equation.
We are thus led to deform the expression~(\ref{fgrad}) of the force $F_\x$ into
\beq
\w F_\x=\sum_{a=1}^D\bigl((1+V_a)S_{\x-\E_a}+(1-V_a)S_{\x+\E_a}\bigr),
\label{fasym}
\eeq
where $V_a$ are arbitrary parameters.
From now on, the vector $\V$ with components~$V_a$ will be interpreted as a velocity.

Asymmetric Langevin dynamics is therefore defined by the linear equation
\beqa
\frac{\d S_\x}{\d t}
&=&\sum_{a=1}^D\bigl((1+V_a)S_{\x-\E_a}+(1-V_a)S_{\x+\E_a}-2S_\x\bigr)\nonumber\\
&-&z(t)S_\x+\eta_\x(t).
\label{lan}
\eeqa

This dynamics is irreversible, i.e., does not obey detailed balance.
It however drives the system to a Gibbsian nonequilibrium stationary state,
where the weights of spin configurations are the Boltzmann weights
associated with the Hamiltonian~(\ref{ham}) at temperature $T$,
irrespective of the velocity~$\V$.
This will be proved in section~\ref{static1}.

\section{One-time observables}
\label{static}

\subsection{Equal-time correlation function}
\label{static1}

From now on, we consider the thermodynamic limit of an infinite system.
We assume that the initial state is the infinite-temperature equilibrium
state with the mean spherical constraint~(\ref{cave}).
The spins $S_\x(0)$ at time $t=0$ are therefore independent and identically
distributed Gaussian variables with unit variance.
The linearity of the Langevin equation~(\ref{lan}) ensures that
the spins $S_\x(t)$ remain Gaussian at any time $t$.
The only non-trivial one-time observable is therefore
the equal-time two-point correlation function
\beq
C_{\x-\y}(t)=\mean{S_\x(t)S_\y(t)},
\label{crt}
\eeq
where brackets denote an average over the infinite-temperature initial state
{\it and} over thermal histories (realizations of the noise).

The main result of this section is as follows.
The equal-time correlation function,
whose explicit expression is given in~(\ref{cq}),
does not depend on $\V$.
It therefore coincides with the equal-time correlation function
for symmetric Langevin dynamics.

Hereafter we only summarize the main properties of this correlation function,
referring the reader to~\cite{gl2D} for more details.
We have
\beq
C_\0(t)=1,
\label{c0t}
\eeq
reflecting the mean spherical constraint~(\ref{cave}), and
\beq
C_\x(0)=\delta_{\x,\0},
\label{cinit}
\eeq
as a consequence of the statistical independence of the spins in the initial state.

In order to pursue with the explicit evaluation of the correlation function $C_\x(t)$,
we have to first solve the Langevin equation~(\ref{lan}) for an infinite system.
This can be readily done in Fourier space.
Define the Fourier transform by the formulas
\beq
f^\F(\q)=\sum_\x f_\x\,\e^{-\ii\q.\x},\qquad
f_\x=\int\dq\,f^\F(\q)\,\e^{\ii\q.\x},
\eeq
where
\beq
\int\dq=\int_{-\pi}^{\pi}\frac{\d q_1}{2\pi}
\cdots\int_{-\pi}^{\pi}\frac{\d q_D}{2\pi}
\eeq
is the normalized integral over the first Brillouin zone.
Equation~(\ref{lan}) thus becomes
\beq
\frac{\dpar S^\F(\q,t)}{\dpar t}=-[\Omega(\q)+z(t)]S^\F(\q,t)+\eta^\F(\q,t),
\label{lanq}
\eeq
where
\beq
\Omega(\q)=2\sum_{a=1}^D(1-\cos q_a+\ii V_a\sin q_a)
\label{Omdef}
\eeq
and
\beq
\mean{\eta^\F(\q,t)\eta^\F(\q',t')}
=2T\,(2\pi)^D\,\delta^D(\q+\q')\,\delta(t-t').
\label{etavar}
\eeq
The solution to~(\ref{lanq}) reads
\beqa
S^\F(\q,t)&=&\e^{-\Omega(\q)t-Z(t)}\nonumber\\
&\times&\left(S^\F(\q,0)+\int_0^t\d s\,\e^{\Omega(\q)s+Z(s)}\eta^\F(\q,s)\right),
\label{sq}
\eeqa
with
\beq
Z(t)=\int_0^t\d s\,z(s).
\eeq

The equal-time correlation function
$C^\F(\q,t)$ in Fourier space is defined by
\beq
\mean{S^\F(\q,t)S^\F(\q',t)}=(2\pi)^D\delta^D(\q+\q')\,C^\F(\q,t).
\label{cqdef}
\eeq
By averaging the expression~(\ref{sq}) over the white noise $\eta^\F(\q,t)$,
whose variance is given by~(\ref{etavar}), and using the initial condition
\beq
C^\F(\q,0)=1
\eeq
implied by~(\ref{cinit}), we obtain the expression
\beq
C^\F(\q,t)=\frac{\e^{-2\omega(\q)t}}{g(t)}
\left(1+2T\!\int_0^t\d s\,\e^{2\omega(\q)s}g(s)\right),
\label{cq}
\eeq
where
\beq
\omega(\q)=\frac12(\Omega(\q)+\Omega(-\q))=2\sum_{a=1}^D(1-\cos q_a)
\label{omdef}
\eeq
and
\beq
g(t)=\e^{2Z(t)}.
\label{gdef}
\eeq

The expression~(\ref{omdef}) for $\omega(\q)$ is independent of the velocity $\V$.
We have thus completed the proof of the result announced above,
namely that the equal-time correlation function~(\ref{cq}) is independent of $\V$.
As a consequence, one-time observables are identical for symmetric (reversible)
and for asymmetric (irreversible) dynamics.
This striking property relies fundamentally on the linearity of the Langevin equation.

\subsection{Relaxation time and phase diagram}
\label{static2}

We proceed with the calculation of the relaxation time of the dynamics.
The Lagrange multiplier $z(t)$, and thereby the function $g(t)$, are yet to be
determined from the constraint~(\ref{c0t}),~i.e.,
\beq
\int\dq\,C^\F(\q,t)=1.
\label{cint}
\eeq
This identity yields a Volterra integral equation~\cite{cug,gl2D} for $g(t)$,
\beq
g(t)=f(t)+2T\!\int_0^t\d s\,f(t-s)g(s),
\label{volterra}
\eeq
with
\beq
f(t)=\int\dq\,\e^{-2\omega(\q)t}=\left(\e^{-4t}I_0(4t)\right)^D
\comport{\approx}{t\to\infty}(8\pi t)^{-D/2},
\label{fdef}
\eeq
where
\beq
I_n(z)=\int\frac{\d q}{2\pi}\,\e^{z\cos q+\ii nq}
\label{idef}
\eeq
is the modified Bessel function.
In Laplace space, (\ref{volterra}) yields
\beq
g^\L(p)=\frac{f^\L(p)}{1-2Tf^\L(p)},
\label{fglap}
\eeq
with
\beq
f^\L(p)=\int_0^\infty\d t\, f(t)\,\e^{-pt}=\int\dq\,\frac{1}{p+2\omega(\q)}.
\label{fldef}
\eeq
The latter function has a branch point at $p=0$,
with a singular part of the form
\beq
f^\L_\sg(p)\comport{\approx}{p\to0}(8\pi)^{-D/2}\Gamma(1-D/2)p^{(D-2)/2},
\label{fsg}
\eeq
as well as a regular part of the form
\beq
f^\L_\reg(p)=A_1-A_2p+A_3p^2+\cdots,
\label{freg}
\eeq
where the integrals
\beq
A_k=\int\dq\,\frac{1}{(2\omega(\q))^k}
\label{adef}
\eeq
are convergent for $D>2k$.
Equations~(\ref{fsg}) and~(\ref{freg}) jointly determine
the small-$p$ behavior of $f^\L(p)$, as a function of dimension~$D$:
\beqa
D<2:&&\ f^\L(p)\approx(8\pi)^{-D/2}\Gamma(1-D/2)p^{-(1-D/2)},\cr
2<D<4:&&\ f^\L(p)\approx A_1-(8\pi)^{-D/2}\abs{\Gamma(1-D/2)}p^{(D-2)/2},\cr
D>4:&&\ f^\L(p)\approx A_1-A_2p.
\label{flist}
\eeqa
Whenever $D=2,4,\dots$ is an even integer,
the regular and singular parts merge, giving rise to logarithmic corrections.

As is well known, {\it in low enough dimension ($D\le2$), the spherical model
has no phase transition.}
In the present framework, the existence of a phase transition is reflected by
the divergence of the relaxation time.
Such a divergence does not occur in low dimension.
Indeed $f^\L(p)$ diverges as $p\to0$.
As a consequence, at any non-zero temperature $T$,
$g^\L(p)$ has a pole at $p=2\mu^2$, where $T$ and $\mu^2$ are related by
\beq
\frac{1}{T}=2f^\L(2\mu^2)=\int\dq\frac{1}{\omega(\q)+\mu^2}.
\label{mudef}
\eeq
We have
\beq
g(t)\sim\e^{2\mu^2t},
\label{ghigh}
\eeq
and so $z(t)$ tends to the constant $z(\infty)=\mu^2$.
The system relaxes exponentially fast to its nonequilibrium stationary state,
with the finite relaxation time\footnote{Throughout this work the subscript `$\eq$' denotes
the value of a quantity in the equilibrium state generated by symmetric dynamics.}
\beq
\tau_\eq=\frac{1}{2\mu^2},
\label{taudef}
\eeq
which is the same as for symmetric dynamics.
The parameter $\mu$ defined by~(\ref{mudef}) will be interpreted
as the inverse correlation length in the continuum limit (see~(\ref{cst})).
The relation~(\ref{taudef}) therefore demonstrates that the dynamical exponent
of the model has the classical value $z=2$,
characteristic of diffusion processes, irrespectively of dimension.

The parameter $\mu$ is an increasing function of temperature,
diverging as
\beq
\mu^2\comport{\approx}{T\to\infty}T-2D+\cdots
\label{muhigh}
\eeq
at high temperature (in any dimension $D$), as can be seen by
expanding~(\ref{mudef}),
and vanishing as $T\to0^+$ for $D\le2$, and as $T\to T_c^+$ for $D>2$ (see below).

On the one-dimensional chain, we have
\beq
f^\L(p)=\frac{1}{\sqrt{p(p+8)}},
\label{flap1}
\eeq
so that the dependence of $\mu^2$ on temperature reads
\beq
T=\mu\sqrt{\mu^2+4},\qquad\hbox{i.e.,}\;\;\mu^2=\sqrt{T^2+4}-2,
\label{Tmu}
\eeq
thus
\beq
\mu\comport{\approx}{T\to0}\frac{T}{2}
\eeq
vanishes linearly at low temperature.

On the two-dimensional square lattice,
\beq
f^\L(p)=\frac{2}{\pi\abs{p+8}}\;\K\!\left(\frac{8}{\abs{p+8}}\right),
\eeq
can be expressed in terms of the complete elliptic integral $\K$.
We have therefore
\beq
\frac{1}{T}=\frac{2}{\pi(\mu^2+4)}\;\K\!\left(\frac{4}{\mu^2+4}\right)
\comport{\approx}{\mu\to0}\frac{1}{4\pi}\ln\frac{32}{\mu^2},
\label{t2k}
\eeq
hence
\beq
\mu\comport{\approx}{T\to0}4\sqrt2\,\e^{-2\pi/T}
\label{mutwo}
\eeq
vanishes exponentially fast at low temperature.

{\it In high enough dimension ($D>2$), the spherical model has a phase transition
at a non-zero critical temperature}
\beq
T_c=\frac{1}{2A_1}=\left(\int\dq\,\frac{1}{\omega(\q)}\right)^{-1}.
\label{tc}
\eeq
Indeed $f^\L(0)=A_1$ is finite,
and so the pole of $g^\L(p)$ hits the tip of the cut of~$f^\L(p)$ at $p=0$
at a finite critical temperature $T_c$.
The numerical values of $T_c$ in various dimensions are given in table~\ref{tcic}
(section~\ref{stat}).

Equations~(\ref{mudef})--(\ref{taudef}) still hold in the paramagnetic phase ($T>T_c$).
The parameter $\mu$ is still an increasing function of temperature,
diverging as~(\ref{muhigh}) at high temperature,
and vanishing as the critical point is approached from above
($T\to T_c^+$), in a dimension-dependent way, according to
\beqa
2<D<4:&&\mu\comport{\approx}{T\to T_c^+}
\left(\frac{(4\pi)^{D/2}\,(T-T_c)}{\abs{\Gamma(1-D/2)}T_c^2}\right)^{1/(D-2)},\cr\cr
D>4:&&\mu\comport{\approx}{T\to
T_c^+}\left(\frac{T-T_c}{4A_2T_c^2}\right)^{1/2}.
\label{muas}
\eeqa
These estimates obey the scaling law $\mu\sim(T-T_c)^\nu$,
where the critical exponent of the correlation length reads
\beqa
2<D<4:&&\ \nu=\frac{1}{D-2},\cr
D>4:&&\ \nu=\frac{1}{2}.
\label{nures}
\eeqa

At the critical point $(T=T_c)$, we have
\beqa
2<D<4:&&\ g(t)\comport{\approx}{t\to\infty}
(D-2)(8\pi)^{(D-2)/2}\sin\left[(D-2)\frac{\pi}{2}\right]\frac{t^{-(4-D)/2}}{T_c^2},\cr
D>4:&&\ g(t)\comport{\to}{t\to\infty}\frac{1}{4A_2T_c^2}.
\label{gcrit}
\eeqa

In the ferromagnetic phase $(T<T_c)$, we have
\beq
g(t)\comport{\approx}{t\to\infty}\frac{f(t)}{M_\eq^4}
\approx\frac{(8\pi t)^{-D/2}}{M_\eq^4},
\label{glow}
\eeq
where the spontaneous magnetization $M_\eq$ is given by~\cite{baxter}
\beq
M_\eq^2=1-\frac{T}{T_c}.
\label{mdef}
\eeq
The function $g(t)$ obeys the sum rule
\beq
\int_0^\infty\d t\, g(t)=\frac{1}{2T_cM_\eq^2}.
\label{fgid}
\eeq

\section{Two-time observables}
\label{two}

In this section, devoted to two-time quantities,
we derive the general expressions~(\ref{cqts}) and~(\ref{rqts})
of the two-time correlation and response functions in Fourier space,
and remind the definition of the so-called \fd ratio~(\ref{xdef}).

\subsection{Correlation function}

The two-time correlation function is defined as
\beq
C_{\x-\y}(t,s)=\mean{S_\x(t)S_\y(s)},
\eeq
with $0\le s$~(waiting time) $\le t$~(observation time).

Its Fourier transform $C^\F(\q,t,s)$ is defined in analogy with~(\ref{cqdef}).
Using~(\ref{sq}), we obtain
\beq
C^\F(\q,t,s)=\frac{\e^{-\Omega(\q)t-\Omega(-\q)s}}{\sqrt{g(t)g(s)}}
\left(1+2T\!\int_0^s\d u\,\e^{2\omega(\q)u}g(u)\right),
\label{cqts1}
\eeq
or equivalently
\beq
C^\F(\q,t,s)=C^\F(\q,s)\,\e^{-\Omega(\q)(t-s)}\sqrt\frac{g(s)}{g(t)},
\label{cqts}
\eeq
where $C^\F(\q,s)$ is given by~(\ref{cq}).

In the following, we shall be mostly interested in the local two-time
correlation function
\beq
C(t,s)=C_\0(t,s)=\mean{S_\x(t)S_\x(s)}=\int\dq\,C^\F(\q,t,s),
\label{cts}
\eeq
also referred to as autocorrelation or diagonal correlation function.

\subsection{Response function}

Suppose now that the system is subjected to a small magnetic field $H_\x(t)$,
depending on the site $\x$ and on time $t\ge0$ in an arbitrary fashion.
This amounts to adding to the ferromagnetic Hamiltonian~(\ref{ham})
a time-dependent perturbation of the form
\beq
\delta\H(t)=-\sum_\x H_\x(t)S_\x(t).
\eeq
Within the present framework,
it is natural to modify the Langevin equation~(\ref{lan})
by adding the magnetic field $H_\x(t)$ to the force $F_\x$ acting on spin $S_\x$,
so that the perturbed equation reads
\beqa
\frac{\d S_\x}{\d t}
&=&\sum_{a=1}^D\bigl((1+V_a)S_{\x-\E_a}+(1-V_a)S_{\x+\E_a}-2S_\x\bigr)\nonumber\\
&-&z(t)S_\x+H_\x(t)+\eta_\x(t).
\label{lanh}
\eeqa
The solution to this equation reads, in Fourier space,
\beqa
S^\F(\q,t)&=&\e^{-\Omega(\q)t-Z(t)}\nonumber\\
&\times&\left(S^\F(\q,0)+\int_0^t\d s\,\e^{\Omega(\q)s+Z(s)}
\left[H^\F(\q,s)+\eta^\F(\q,s)\right]\right).\nonumber\\
\label{lanhsol}
\eeqa
It can indeed be checked that the Lagrange multiplier $\lambda(t)$
remains unchanged, to first order in the magnetic field.

Causality and invariance under spatial translations imply that we have,
to first order in the magnetic field $H_\x(t)$,
\beq
\mean{S_\x(t)}\approx\int_0^t\d s\sum_\y R_{\x-\y}(t,s)H_\y(s),
\label{sxt}
\eeq
where
\beq
R_{\x-\y}(t,s)=\left.\frac{\delta\mean{S_\x(t)}}
{\delta H_\y(s)}\right\vert_{\{H_\x(t)=0\}}
\eeq
is the two-time response function of the model.
As a consequence of~(\ref{lanhsol}), this response function reads, in Fourier
space
\beq
R^\F(\q,t,s)=\e^{-\Omega(\q)(t-s)}\sqrt\frac{g(s)}{g(t)}.
\label{rqts}
\eeq
By~(\ref{cqts}) and~(\ref{rqts}) we obtain the identity
\beq
C^\F(\q,t,s)=C^\F(\q,s)R^\F(\q,t,s)
\label{combi}
\eeq
between the correlation and response functions.
Here again, we shall be mostly inte\-rested in the local two-time response function
\beq
R(t,s)=R_\0(t,s)
=\left.\frac{\delta\mean{S_\x(t)}}{\delta H_\x(s)}\right\vert_{\{H_\x(t)=0\}}
=\int\dq\,R^\F(\q,t,s),
\label{rts}
\eeq
also referred to as autoresponse or diagonal response function.

\subsection{Fluctuation-dissipation ratio}

In an arbitrary stationary state (not necessarily an equilibrium one),
time translational invariance implies that
two-time quantities only depend on the time difference $\tau=t-s$.
We thus have in particular\footnote{Throughout this work the subscript `$\st$' denotes
the value of a quantity in the nonequilibrium stationary state.}
\beq
C_\st(\tau)=\lim_{s\to\infty}C(s+\tau,s),\quad
R_\st(\tau)=\lim_{s\to\infty}R(s+\tau,s).
\eeq

Consider now an equilibrium state,
i.e., a stationary state for a reversible dyn\-amics.
In such a circumstance, the equilibrium correlation and response functions,
denoted as $C_\eq(\tau)$ and $R_\eq(\tau)$,
are related to each other by the \fd theorem~\cite{kubo}
(see~\cite{chandler} for a general introduction):
\beq
T\,R_\eq(\tau)=-\frac{\d C_\eq(\tau)}{\d\tau}.
\label{fdt}
\eeq

One possible characterization of the distance to equilibrium is the so-called
\fd ratio~\cite{fd}
(see~\cite{fdrev,calab} for reviews):
\beq
X(t,s)=\frac{T\,R(t,s)}{\,\frad{\dpar C(t,s)}{\dpar s}\,}.
\label{xdef}
\eeq

In a nonequilibrium stationary state, the \fd theorem does not hold in general,
and so the \fd ratio remains a non-trivial function
\beq
X_\st(\tau)=\lim_{s\to\infty}X(s+\tau,s).
\label{xstdef}
\eeq

Likewise, in order to characterize the transient regime before the stationary
state sets in, we consider the asymptotic \fd ratio
\beq
X_\as(s)=\lim_{\tau\to\infty}X(s+\tau,s).
\label{xasdef}
\eeq

\section{Nonequilibrium stationary state: general properties}
\label{stat}

This section as well as the next two ones are devoted to the study of the nonequilibrium stationary state
attained by the system in its paramagnetic phase, i.e., for $T>T_c$.
We recall that the critical temperature $T_c$ is non-zero for $D>2$ only.
In this section we study general properties of the nonequilibrium stationary state,
whereas more specific features will be successively discussed
in section~\ref{stat1} in the one-dimensional case
and in section~\ref{stathigh} in higher dimension.

\subsection{Dispersive \fd ratio}

In the stationary state,
the function $g(t)$ grows exponentially as $\e^{2\mu^2t}$
(see~(\ref{ghigh})), where $\mu^2$ is given by~(\ref{mudef}).
The expression~(\ref{cq}) for the equal-time correlation function in Fourier space
thus yields the static structure factor
\beq
C_\eq^\F(\q)=\frac{T}{\omega(\q)+\mu^2}.
\label{cst}
\eeq
Equations~(\ref{cqts}) and~(\ref{rqts}) then simplify to
\beqa
C_\st^\F(\q,\tau)&=&\frac{T}{\omega(\q)+\mu^2}\,\e^{-(\Omega(\q)+\mu^2)\tau},
\nonumber\\
R_\st^\F(\q,\tau)&=&\e^{-(\Omega(\q)+\mu^2)\tau}.
\label{crst}
\eeqa
These expressions fulfil the stationary form of the identity~(\ref{combi}), i.e.,
\beq
C_\st^\F(\q,\tau)=C_\eq^\F(\q)R_\st^\F(\q,\tau).
\label{combist}
\eeq
We thus obtain a generalized \fd relation in Fourier space, of the form
\beq
T\,R_\st^\F(\q,\tau)=-X_\st(\q)\,\frac{\d C_\st^\F(\q,\tau)}{\d\tau},
\eeq
where the \fd ratio
\beq
X_\st(\q)=\frac{\omega(\q)+\mu^2}{\Omega(\q)+\mu^2}
\label{xq}
\eeq
(see~(\ref{Omdef}),~(\ref{omdef}))
is complex and dispersive, i.e., $\q$-dependent, but independent of $\tau$\footnote{A \fd relation in momentum space was already considered in~\cite{calab}.}.

At the critical point ($T=T_c$),
all the above results still hold, albeit with $\mu=0$.
In the ferromagnetic phase ($T<T_c$),
the static structure factor includes a delta peak at $\q=\0$,
proportional to the square spontaneous magnetization $M_\eq^2$, i.e.,
\beq
C_\eq^\F(\q)=\frac{T}{\omega(\q)}+M_\eq^2\,(2\pi)^D\delta^D(\q).
\label{cstferro}
\eeq
The expression~(\ref{mdef}) of $M_\eq^2$ can be recovered by
integrating~(\ref{cstferro}) over $\q$, using~(\ref{cint}) and~(\ref{tc}).

\subsection{Entropy production rate}
\label{epr}

The evaluation of the entropy production rate
in the nonequilibrium stationary state of the spherical model
with asymmetric Langevin dynamics was the subject of~\cite{entropy}.
With the notations of the present work, the result
for the stationary entropy production rate~$\phi$ per spin and per unit time reads
\beq
\phi=\frac{1}{T}\bigmean{\w F_\x\bigl(\w F_\x-F_\x\bigr)},
\eeq
where $F_\x$ and $\w F_\x$ are the forces acting on spin $S_\x$
for symmetric dynamics (see~(\ref{fgrad}))
and for asymmetric dynamics (see~(\ref{fasym})), respectively.
Using the symmetries of the hypercubic lattice, one can recast
the above expression into the following one,
\beq
\phi=\frac{1}{T}\bigmean{\bigl(\w F_\x-F_\x\bigr)^2},
\eeq
which is more appealing, as it is manifestly positive.

The above expression can be further reduced to
\beq
\phi=\frac{2}{T}(1-C_{2,\eq})\V^2,
\label{phi}
\eeq
where
\beq
C_{2,\eq}=T\!\int\dq\frac{\cos 2q_1}{\omega(\q)+\mu^2}
\eeq
is the static (i.e., equilibrium) correlation function
two sites apart along any axis of the lattice (e.g.~$\x=2\E_1$).
The stationary entropy production rate
only depends on temperature and on the square of the velocity vector.

At the critical point ($T=T_c$, i.e., $\mu=0$),
the expression~(\ref{phi}) becomes
\beq
\phi_c=2\,I_c\V^2,
\label{phic}
\eeq
where
\beq
I_c=\int\dq\frac{1-\cos 2q_1}{\omega(\q)}.
\eeq
The result~(\ref{phic}) also holds for $D=1$ and $D=2$ at $T=T_c=0$.
Table~\ref{tcic} gives the values of the critical temperature $T_c$ (see~(\ref{tc}))
and of the integral $I_c$ for various values of the spatial dimension $D$.

\begin{table}[!ht]
\begin{center}
\begin{tabular}{|c||c|c|}
\hline
$D$ & $T_c$ & $I_c$ \\
\hline
1 & 0 & 1 \\
2 & 0 & $1-\frac{2}{\pi}\approx0.363380$ \\
3 & 3.956776 & 0.209841 \\
4 & 6.454386 & 0.146687 \\
\hline
$D\gg1$ & $2D$ & $1/(2D)$ \\
\hline
\end{tabular}
\end{center}
\caption{\small Numerical values of the critical temperature $T_c$
of the model (see~(\ref{tc}))
and of the integral $I_c$ entering the expression~(\ref{phic})
of the entropy production rate at $T_c$,
for the first few integer values of dimension~$D$.
Last line: leading asymptotic behavior at large $D$.}
\label{tcic}
\end{table}

The entropy production rate remains constant and equal to $\phi_c$ in the whole
ferromagnetic phase ($T\le T_c$).
This property is a consequence of~(\ref{cstferro}),
which implies that the difference $1-C_{\x,\eq}$ is proportional to
$1-M_\eq^2=T/T_c$ for all temperatures $T\le T_c$.
The entropy production rate then decreases monotonically from $\phi_c$ to zero
in the paramagnetic phase ($T\ge T_c$).
At high temperature, we have $\phi\approx2\V^2/T$,
since $C_{2,\eq}$ becomes negligible, as it falls off as $1/T^2$.

\section{Nonequilibrium stationary state: one dimension}
\label{stat1}

Let us now discuss specific features of the nonequilibrium stationary state
on the one-dimensional chain.
The expression~(\ref{cst}) of the static structure factor reads
\beq
C_\eq^\F(q)=\frac{T}{2(1-\cos q)+\mu^2}.
\eeq
Using the identity
\beq
\sum_{x=-\infty}^{+\infty}\e^{-\m\abs{x}-\ii qx}=\frac{\sinh\m}{\cosh\m-\cos q},
\eeq
we obtain the static correlation function
\beq
C_{x,\eq}=\mean{S_xS_0}=\e^{-\m\abs{x}},
\label{cstx}
\eeq
where the inverse equilibrium correlation length
\beq
\m=\frac{1}{\xi_\eq}
\eeq
is related to $\mu$ and $T$ by
\beq
\mu=2\sinh\frac{\m}{2},\qquad T=2\sinh\m
\label{mt}
\eeq
(see~(\ref{Tmu})).
The expression~(\ref{phi}) for the entropy production rate simplifies~to
\beq
\phi=2\,\e^{-\m}\,V^2.
\eeq

In the stationary state the expressions~(\ref{cts}) and~(\ref{rts}) for
the correlation and response functions become
\beqa
C_\st(\tau)
&=&T\,\e^{-(2+\mu^2)\tau}\int\frac{\d q}{2\pi}
\frac{\e^{2(\cos q-\ii V\sin q)\tau}}{2(1-\cos q)+\mu^2}\nonumber\\
&=&T\!\int_0^\infty\d u\,\e^{-(2+\mu^2)(\tau+u)}
I_0\left(2\sqrt{(\tau+u)^2-V^2\tau^2}\right),
\label{c1}
\\
R_\st(\tau)
&=&\e^{-(2+\mu^2)\tau}\int\frac{\d q}{2\pi}
\,\e^{2(\cos q-\ii V\sin q)\tau}\nonumber\\
&=&\e^{-(2+\mu^2)\tau}I_0\left(2\sqrt{1-V^2}\,\tau\right),
\label{r1}
\eeqa
where $I_0$ is the modified Bessel function~(see~(\ref{idef})).

These expressions call for the following comment.
The velocity $V$, introduced as a deformation parameter in~(\ref{fasym}),
may a priori take arbitrary values.
As already mentioned, we can restrict ourselves to the case where $V$ is positive.
The value $V=1$ turns out to play a special role,
demarcating a usual regime ($V<1$),
where the correlation and response functions
are positive and decay monotonically to zero,
and an unusual one ($V>1$),
where the correlation and response functions
exhibit oscillations as a function of time.
The same phenomenon appears in the asymmetric dynamics of the Ising chain~\cite{cgtocome}.

\subsection{Low-temperature scaling regime}
\label{stat1low}

We first consider the scaling regime
where the temperature $T\approx2\mu\approx2\m$ and the velocity $V$ are
simultaneously small.
In this regime, it is legitimate to treat $\mu$, $V$ and~$q$ on the same footing,
and to consistently use $\omega(q)\approx q^2$ and $\Omega(q)\approx q^2+2\ii Vq$.
This approximation amounts to replacing
lattice propagators (given by products of Bessel functions)
by the much simpler continuum ones (given by Gaussian functions).
In the following we shall meet other instances of scaling regimes
which lend themselves to such a continuum treatment.

Within this framework, the expressions~(\ref{c1}), (\ref{r1})
for the correlation and response functions simplify to
\beqa
C_\st(\tau)&\approx&\frac12
\Bigl[\e^{2\mu V\tau}\erfc\left((\mu+V)\sqrt\tau\right)
+\e^{-2\mu V\tau}\erfc\left((\mu-V)\sqrt\tau\right)\Bigr],\nonumber\\
R_\st(\tau)&\approx&\frac{\e^{-(\mu^2+V^2)\tau}}{\sqrt{4\pi\tau}},
\label{crlow}
\eeqa
where
\beq
\erf z=\frac{2}{\sqrt\pi}\int_0^z\d y\,\e^{-y^2},\qquad\erfc z=1-\erf z
\label{erfdef}
\eeq
are respectively the error function and the complementary error function.
The ex\-pres\-sions~(\ref{crlow}) are even functions of $V$, as it should be,
so that it is sufficient to consider the case where $V\ge0$.

The corresponding \fd ratio $X_\st(\tau)$ is a non-trivial function of~$\tau$,
as expected in a nonequilibrium stationary state, which is given by
\beqa
\frac{1}{X_\st(\tau)}\approx1&+&V\sqrt{\pi\tau}\,\e^{(\mu^2+V^2)\tau}\nonumber\\
&\times&\Bigl[\e^{-2\mu V\tau}\!\erfc\left((\mu-V)\sqrt\tau\right)
-\e^{2\mu V\tau}\!\erfc\left((\mu+V)\sqrt\tau\right)\Bigr].\nonumber\\
\label{xlow}
\eeqa
The long-time behavior of this expression
depends on the relative values of $V$ and~$\mu$.

\begin{itemize}

\item{\it Regime~I ($V<\mu$).}
In this regime, the correlation function falls off as
\beq
C_\st(\tau)\approx\frac{2\mu}{\mu^2-V^2}\,\frac{\e^{-(\mu^2+V^2)\tau}}{\sqrt{4\pi\tau}},
\eeq
while the response function is given by~(\ref{crlow}).
The decay rates $\alpha_C$ and $\alpha_R$ of the correlation and of the response
are therefore equal:
\beq
\alpha_C=\alpha_R=\mu^2+V^2,
\eeq
and the \fd ratio has a finite asymptotic value
\beq
X_\st(\infty)=\frac{\mu^2-V^2}{\mu^2+V^2},
\label{xinfstat}
\eeq
which decreases continuously from unity as $V$ increases, and vanishes as $V\to\mu$.
Regime~I is therefore characterized by a {\it weak violation of the \fd theorem}.

\item{\it Regime~II ($V>\mu$).}
In this regime, the correlation function falls off as
\beq
C_\st(\tau)\approx\e^{-2\mu V\tau},
\eeq
while the response function is still given by~(\ref{crlow}).
The decay rates $\alpha_C$ and $\alpha_R$ are now different:
\beq
\alpha_C=2\mu V<\alpha_R=\mu^2+V^2,
\eeq
hence the \fd ratio decays exponentially fast,
with a decay rate $\alpha_R-\alpha_C=(V-\mu)^2$, and so
\beq
X_\st(\infty)=0.
\eeq
Regime~II is therefore characterized by a {\it strong violation of the \fd theorem}.

\item{\it Borderline situation ($V=\mu$).}
In this case, the correlation function falls off~as
\beq
C_\st(\tau)\approx\frac{\e^{-2\mu^2\tau}}{2},
\eeq
and so $\alpha_C=\alpha_R=2\mu^2$.
The \fd ratio however decays very slowly, as
\beq
X_\st(\tau)\approx\frac{1}{\mu\sqrt{\pi\tau}}.
\eeq

\end{itemize}

\begin{figure}[!ht]
\begin{center}
\includegraphics[angle=0,width=0.7\linewidth]{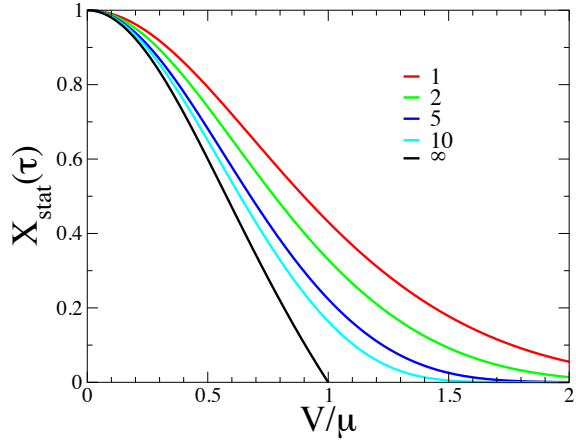}
\caption{\small
Plot of the asymptotic \fd ratio $X_\st(\tau)$ pertaining
to the nonequilibrium stationary state on the chain
in the low-temperature regime (see~(\ref{xlow})),
against the ratio $V/\mu$, for $\mu^2\tau=1$, 2, 5,~10 and $\infty$
(the latter case corresponding to the asymptotic result~(\ref{xinfstat})).}
\label{xplot}
\end{center}
\end{figure}

The above discussion is very reminiscent of the case
of the Ising chain under asymmetric dynamics~\cite{cg2011}.
It is summarized in figure~\ref{xplot},
showing a plot of $X_\st(\tau)$ against the ratio $V/\mu$,
for several finite values of the dimensionless product $\mu^2\tau$,
as well as the asymptotic result~(\ref{xinfstat}).

\subsection{Arbitrary finite temperature}

We now turn to the analysis of the nonequilibrium stationary state on the chain
for an arbitrary finite temperature~$T$.

The response function $R_\st(\tau)$ falls off exponentially
as $R_\st(\tau)\sim\e^{-\alpha_R\tau}$, with
\beq
\alpha_R=2\left(1-\sqrt{1-V^2}\right)+\mu^2.
\label{ar}
\eeq
This asymptotic result can be derived in two ways:
either by using the second line of~(\ref{r1})
and the asymptotic growth $I_0(z)\sim\e^z$ of the modified Bessel function,
or by evaluating the integral in the first line of~(\ref{r1})
by the saddle-point method.
The saddle-point equation reads $\tan q=-\ii V$.

The asymptotic analysis of the correlation function $C_\st(\tau)$ is more intricate.
Consider the integral in the first line of~(\ref{c1}).
Its asymptotic behavior at large $\tau$ can be a priori
governed by two special values of $q$:
the saddle point at $q\un$ such that $\tan q\un=-\ii V$, yielding the decay
rate
\beq
\alpha\un=\alpha_R
\eeq
(see~(\ref{ar})),
and the zero of the denominator at $q\de$ such that $\cos q\de=1+\mu^2/2$,
yielding the decay rate
\beq
\alpha\de=V\mu\sqrt{\mu^2+4}=VT.
\label{a2}
\eeq
Both $q\un$ and $q\de$ turn out to be purely imaginary.
Hence the relevant decay rate is that pertaining to the special value of $q$
which is closer to the real axis, i.e., whose absolute imaginary part is the smaller.
We have $q\un=q\de$ (and so $\alpha\un=\alpha\de$) at the borderline velocity
\beq
V_\B=\frac{\mu\sqrt{\mu^2+4}}{\mu^2+2}=\frac{T}{\sqrt{T^2+4}}=\tanh\m.
\label{v0def}
\eeq
The scenario put forward in the low-temperature scaling regime extends
to the stationary state at any finite temperature.

\begin{itemize}

\item{\it Regime~I ($V<V_\B$).}
In this regime, $q\un$ is closer to the real axis than $q\de$, hence
\beq
\alpha_C=\alpha_R=\alpha\un.
\label{a1}
\eeq
The asymptotic \fd ratio
\beq
X_\st(\infty)
=\frac{2\left(1-1\left/\sqrt{1-V^2}\right.\right)+\mu^2}
{2\left(1-\sqrt{1-V^2}\right)+\mu^2},
\label{xinf1}
\eeq
i.e.,
\beq\label{eq:xstat}
X_{\st}(\infty)=
\frac{1-\sqrt{1-V_\B^2}\left/\sqrt{1-V^2}\right.}{1-\sqrt{1-V_\B^2}\sqrt{1-V^2}},
\eeq
is consistently equal to the value of~(\ref{xq}) at the saddle point ($q=q\un$).
It decreases monotonically from unity to zero
as $V$ is increased from 0 to~$V_\B$.
Let us note that the expression~(\ref{eq:xstat}) has exactly the same form as in the case of the Ising chain with asymmetric dynamics~\cite{cg2011}.
In the latter case, the relationship between $V_\B$ and temperature is given by
$V_\B=\sqrt{1-\gamma^2}$, where $\gamma=\tanh 2/T$.
As in~(\ref{v0def}), $V_\B$ increases from $0$ to $1$, when $T$ varies between $0$ and infinity.

\item{\it Regime~II ($V>V_\B$).}
Now $q\de$ is closer to the real axis than $q\un$, and so we have
\beq
\alpha_C=\alpha\de<\alpha_R=\alpha\un.
\eeq
The \fd ratio falls off exponentially fast to zero,
with a decay rate $\alpha\un-\alpha\de$.

\item{\it Borderline situation ($V=V_\B$).}
We have then
\beq
\alpha_C=\alpha_R=\frac{2 V_\B^2}{\sqrt{1-V_\B^2}}
=\frac{\mu^2(\mu^2+4)}{\mu^2+2}
=\frac{T^2}{\sqrt{T^2+4}}.
\label{bor1}
\eeq

\end{itemize}

\subsection{Spatial interpretation of the two regimes of \fd violation}

The existence of two regimes of \fd violation,
i.e., Regime~I of weak violation, with a finite $X_\st(\infty)$,
and Regime~II of strong violation, with a vanishing $X_\st(\infty)$,
has the following real-space interpretation.
Let us start from the identity~(\ref{combist}),
which translates in real space into the convolution formula
\beq
C_{x,\st}(\tau)=\sum_{y=-\infty}^{+\infty} C_{x-y,\eq}\, R_{y,\st}(\tau).
\label{convol}
\eeq
The static correlation function $C_{x,\eq}$ is given by~(\ref{cstx}),
while the space-dependent response function $R_{x,\st}(\tau)$ reads
\beqa
R_{x,\st}(\tau)
&=&\e^{-(2+\mu^2)\tau}\int\frac{\d q}{2\pi}
\,\e^{\ii qx+2(\cos q-\ii V\sin q)\tau}\nonumber\\
&=&\e^{-(2+\mu^2)\tau}\left(\frac{1+V}{1-V}\right)^{x/2}
I_x\left(2\sqrt{1-V^2}\,\tau\right),
\label{r1x}
\eeqa
where $I_x$ is the modified Bessel function~(see~(\ref{idef})).
A very similar expression of the stationary response is found for the
Ising chain under asymmetric dynamics~\cite{cg2011}.
Consider the expression~(\ref{convol}) for the autocorrelation function
($x=0$), i.e.,
\beq
C_\st(\tau)=\sum_{y=-\infty}^{+\infty} C_{-y,\eq}\, R_{y,\st}(\tau).
\label{convol0}
\eeq
The first factor falls off exponentially as $C_{-y,\eq}=\e^{-\m\abs{y}}$
(see~(\ref{cstx})),
where the decay rate $\m$ is the inverse equilibrium correlation length.
The second factor, $R_{y,\st}(\tau)$, exhibits, in the regime of long times,
an exponential profile of the form
\beq
\frac{R_{y,\st}(\tau)}{R_{0,\st}(\tau)}\approx\e^{\n y},
\eeq
characterized by the growth rate $\n$ or, equivalently,
by the asymmetry length $\xi_\asym$:
\beq
\n=\frac{1}{2}\ln\frac{1+V}{1-V}=\frac{1}{\xi_\asym}.
\eeq
The two spatial rates $\m$ and $\n$,
or, equivalently, the two lengths $\xi_\eq$ and $\xi_\asym$,
coincide for $V=\tanh\m=V_\B$ (see~(\ref{v0def})).

In Regime~I, i.e., for $V<V_\B$, one has $\n<\m$ (i.e., $\xi_\eq<\xi_\asym$).
The product $C_{-y,\eq}\,R_{y,\st}(\tau)$ is therefore maximum at the origin ($y=0$),
and so the sum over sites~$y$ entering~(\ref{convol0}) is dominated
by values of $y$ around the origin.
We thus recover the result $\alpha_C=\alpha_R$ (see~(\ref{a1})).

In Regime~II, i.e., for $V>V_\B$, one has $\n>\m$ (i.e., $\xi_\eq>\xi_\asym$),
and so the sum entering~(\ref{convol0}) is dominated by distant values of $y$.
A more precise statement is as follows.
The asymptotic behavior of the modified Bessel function
\beq
I_n(z)\sim\exp\left(\sqrt{n^2+z^2}-n\ln\frac{n+\sqrt{n^2+z^2}}{z}\right),
\eeq
which holds whenever $n$ and $z$ are simultaneously large,
translates into the asymptotic fall-off of the response function
\beq
R_{x,\st}(\tau)\sim\e^{-\Psi(w)\,\tau}
\label{rasy}
\eeq
in any direction of the $x$--$\tau$ plane such that $x=w\tau$,
where
\beqa
\Psi(w)=2&+&\mu^2-\sqrt{w^2+4(1-V^2)}
\nonumber\\
&+&w\ln\frac{w+\sqrt{w^2+4(1-V^2)}}{2(1+V)}.
\eeqa
The large-deviation function $\Psi(w)$
consistently obeys $\Psi(0)=\alpha_R$ (see~(\ref{ar})),
as well as the symmetry property $\Psi(w)-\Psi(-w)=-2\n w$.
It takes its minimum $\Psi=\mu^2$ for $w=2V$,
so that the response function is maximum near the ballistic point $x=2V\tau$,
where it falls off as $\e^{-\mu^2\tau}$.
The asymptotic decay rate of the maximum response is thus identical
to that of the local response for symmetric dynamics (see~(\ref{ar})).

Inserting the estimate~(\ref{rasy}) into~(\ref{convol0}),
we are left with the expression
\beq
\alpha_C=\min_w\bigl(\Psi(w)-\m w\bigr),
\eeq
where the minimum has to be attained for a positive value of $w$.
We thus obtain
\beq
w=2(V\cosh\m-\sinh\m),\qquad\alpha_C=2V\sinh\m.
\eeq
The first expression yields a positive solution ($w>0$) for $V>V_\B=\tanh\m$,
whereas the second expression is in agreement with~(\ref{mt}),~(\ref{a2}).

\section{Nonequilibrium stationary state: higher dimension}
\label{stathigh}

We now turn to the analysis of specific features
of the nonequilibrium stationary state in higher dimension.

The expressions~(\ref{cts}) and~(\ref{rts}) for
the stationary correlation and response functions read explicitly
\beqa
C_\st(\tau)
&=&\int\dq\frac{T}{\omega(\q)+\mu^2}\,\e^{-(\Omega(\q)+\mu^2)\tau}\nonumber\\
&=&T\!\int_0^\infty\d u\,\e^{-(2D+\mu^2)(\tau+u)}
\prod_{a=1}^D I_0\left(2\sqrt{(\tau+u)^2-V_a^2\tau^2}\right),
\nonumber\\
\label{chigh}
\\
R_\st(\tau)
&=&\int\dq\e^{-(\Omega(\q)+\mu^2)\tau}\nonumber\\
&=&\e^{-(2D+\mu^2)\tau}\prod_{a=1}^D I_0\left(2\sqrt{1-V_a^2}\,\tau\right),
\label{rhigh}
\eeqa
where $I_0$ is the modified Bessel function~(see~(\ref{idef})).

These expressions again demonstrate that the usual situation,
where the cor\-re\-la\-tion and response functions
are positive and decay monotonically to zero,
corresponds to the case where all the components of the velocity obey
$\abs{V_a}<1$.
The asymptotic behavior of $C_\st(\tau)$ and $R_\st(\tau)$
can be studied along the lines of the one-dimensional case (see~section~\ref{stat1}).
The asymptotic analysis of $C_\st(\tau)$
again reveals the existence of two regimes.

\begin{itemize}

\item{\it Regime~I ($\V$ small).}
In this regime, the decay of $C_\st(\tau)$ and $R_\st(\tau)$,
as given by the integral representations
in the first lines of~(\ref{chigh}) and~(\ref{rhigh}),
is dominated by the same saddle point $\q\un$,
defined by $\tan q\un_a=-\ii V_a$.
We have therefore
\beq
\alpha_C=\alpha_R=\alpha\un,
\label{a1high}
\eeq
with
\beq
\alpha\un=2\sum_{a=1}^D\left(1-\sqrt{1-V_a^2}\right)+\mu^2.
\label{arhigh}
\eeq

The asymptotic \fd ratio is given by the value of~(\ref{xq}) at the saddle
point $\q\un$, i.e.,
\beqa
X_\st(\infty)=\frad{2\sum_{a=1}^D\left(1-1\!\left/\!\sqrt{1-V_a^2}\right.\right)+\mu^2}
{2\sum_{a=1}^D\left(1-\sqrt{1-V_a^2}\right)+\mu^2}.
\label{xhigh}
\eeqa
This formula is a generalization of the one-dimensional result~(\ref{xinf1}).

\item{\it Regime~II ($\V$ large).}
In this regime, the decay of $C_\st(\tau)$ is obtained by constrai\-ning the
saddle point to belong to the manifold with equation $\omega(\q)+\mu^2=0$.
In other words, this decay is dominated by the wavevector $\q\de$
so that $\Omega(\q\de)$ is extremal while $\omega(\q\de)+\mu^2$ vanishes.
We have again
\beq
\alpha_C=\alpha\de<\alpha_R=\alpha\un,
\eeq
and the \fd ratio falls off exponentially fast.

More explicitly,
introducing a Lagrange multiplier $\ell$,
we are led to extremalize the quantity
\beq
Q(\q,\ell)=\Omega(\q)-\ell\omega(\q).
\eeq
Introducing for convenience the parameter $L=1/(1-\ell)$,
this amounts to extremalizing
\beq
U(\q,L)=\sum_{a=1}^D(1-\cos q_a+\ii LV_a\sin q_a).
\eeq
The stationarity equations read $\tan q\de_a=-\ii LV_a$.
We thus obtain
\beq
\alpha\de=2L\sum_{a=1}^D\frac{V_a^2}{\sqrt{1-L^2V_a^2}},
\label{a2l}
\eeq
where the parameter $L$ obeys the consistency condition
$\omega(\q\de)+\mu^2=0$, i.e.,
\beq
\mu^2=2\sum_{a=1}^D\left(\frac{1}{\sqrt{1-L^2V_a^2}}-1\right).
\label{mu2l}
\eeq
Regime~II holds as long as $L<1$,
so that the imaginary parts of the $q_a\de$ are smaller than those of the $q_a\un$.

\item{\it Borderline situation.}
The borderline situation between Regimes~I and~II corresponds to $L=1$.
We thus have
\beq
\mu^2=2\sum_{a=1}^D\left(\frac{1}{\sqrt{1-V_a^2}}-1\right).
\label{bs}
\eeq
The numerator of the asymptotic \fd ratio~(\ref{xhigh}) consistently vanishes.
At the borderline, we have
\beq
\alpha_C=\alpha_R=\alpha\un=\alpha\de=2\sum_{a=1}^D\frac{V_a^2}{\sqrt{1-V_a^2}}.
\eeq
This formula is a generalization of the one-dimensional result~(\ref{bor1}).

\end{itemize}

The borderline velocities obeying~(\ref{bs})
run over a closed {\it borderline surface}~$\SS$
in the $D$-dimensional space of velocities $\V$.
Regime~I (weak violation of the \fd theorem),
with its finite asymptotic \fd ratio~(\ref{xhigh}),
holds for velocities inside the surface $\SS$,
while Regime~II (strong violation of the \fd theorem)
holds outside the surface $\SS$.
The size of the borderline surface $\SS$ increases with $\mu$,
i.e., with the distance to the critical point.

\begin{itemize}

\item
For $\mu$ small, i.e., $T$ just above $T_c$,
isotropy in velocity space is restored,
since~$\V$ is only involved through its norm $\abs{\V}$.
In Regime~I, (\ref{arhigh}) and~(\ref{xhigh}) indeed respectively become
\beq
\alpha\un\approx\mu^2+\V^2
\eeq
and
\beq
X_\st(\infty)\approx\frac{\mu^2-\V^2}{\mu^2+\V^2}.
\label{xinfscahigh}
\eeq
The surface $\SS$ is thus a small sphere of radius $\mu$.
In Regime~II, (\ref{a2l}) and~(\ref{mu2l}) yield
\beq
L=\frac{\mu}{\abs{\V}}
\eeq
and
\beq
\alpha\de=2\mu\abs{\V}.
\eeq

\item
In the opposite limit ($\mu$ large, i.e., $T$ large),
the surface $\SS$ approaches the unit cube from inside.

\end{itemize}

\begin{figure}[!ht]
\begin{center}
\includegraphics[angle=-90,width=.55\linewidth]{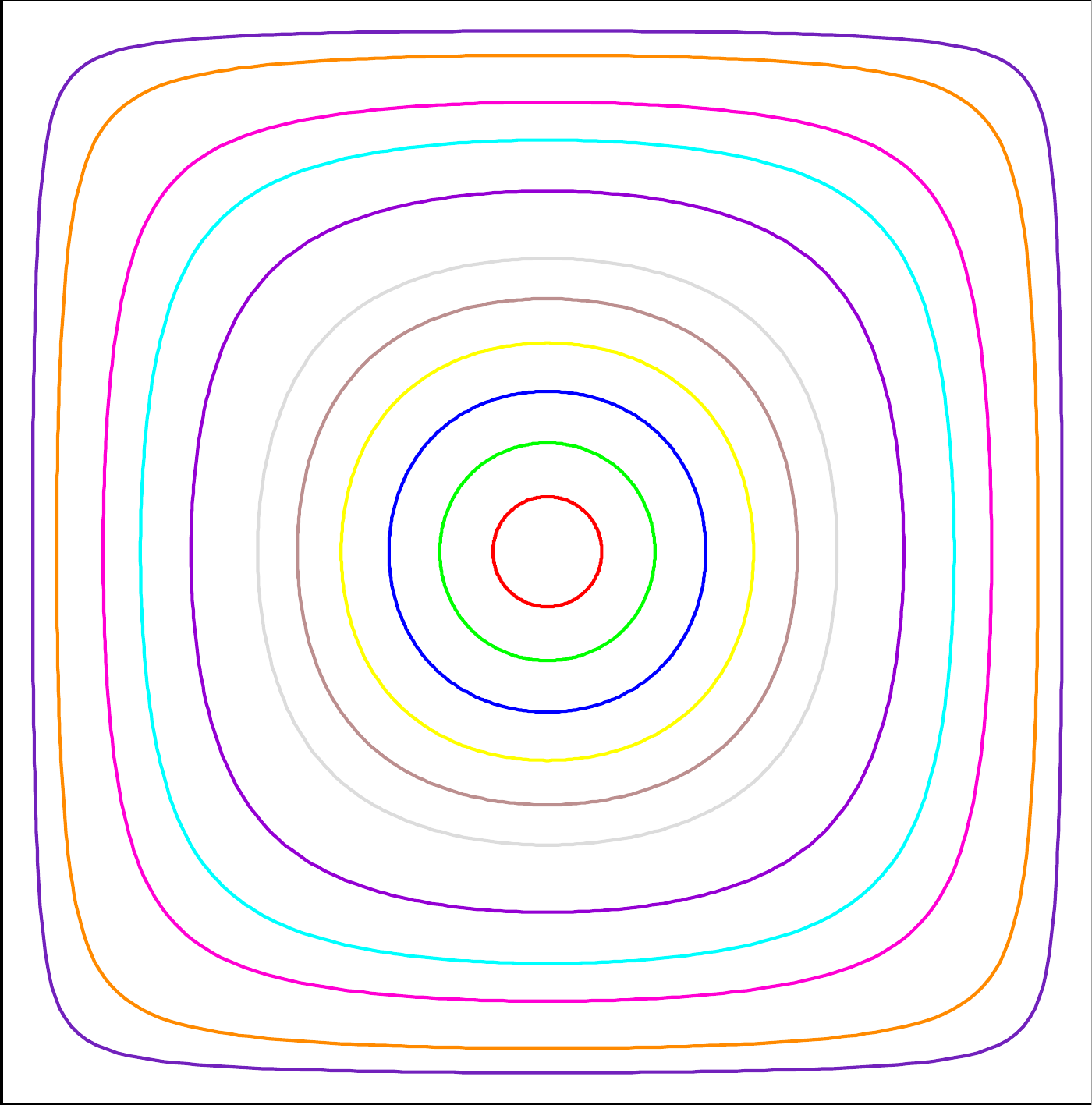}
\caption{\small
Plot of the borderline curve $\SS$ for the square lattice
in the $V_1$--$V_2$ plane, for the values of $\mu$ listed in table~\ref{t2}.
The outermost square is the unit square.}
\label{bl}
\end{center}
\end{figure}

Figure~\ref{bl} shows a plot of the borderline curve $\SS$ for the square lattice
in the $V_1$--$V_2$ plane, with equation
\beq
\frac{1}{\sqrt{1-V_1^2}}+\frac{1}{\sqrt{1-V_2^2}}=2+\frac{\mu^2}{2}
\eeq
(see~(\ref{bs})),
for the values of $\mu$ given in table~\ref{t2},
together with the corresponding temperatures.

\tabcolsep=3pt
\begin{table}[!ht]
\begin{center}
\begin{tabular}{|c||c|c|c|c|c|c|c|c|c|c|c|c|}
\hline
$\mu$ & 0.1 & 0.2 & 0.3 & 0.4 & 0.5 & 0.6 & 0.8 & 1 & 1.2 & 1.6 & 2\\
\hline
$T$&1.559&1.888&2.159&2.410&2.654&2.897&3.399&3.936&4.521&5.860&7.454\\
\hline
\end{tabular}
\end{center}
\caption{\small List of the values of the parameter $\mu$
for which the borderline curve $\SS$ for the square lattice
in shown in figure~\ref{bl},
together with the corresponding values of temperature $T$
(see~(\ref{t2k})).}
\label{t2}
\end{table}

\section{Low-temperature scaling regime in one dimension}
\label{t1}

This section and the three following ones are devoted to the transient regime of
the model, before the nonequilibrium stationary state sets in.
The duration of this regime is measured by the relaxation time
$\tau_\eq=1/(2\mu^2)$ (see~(\ref{taudef})).
The latter is finite in the paramagnetic phase ($T>T_c$),
whereas the transient regime lasts forever
(the system is aging) both at the critical point ($T=T_c$)
and in the whole low-temperature ferromagnetic phase ($T<T_c$).

In this section we consider the scaling behavior
of the one-dimensional model at low temperature,
in the regime where $\mu$ and $V$ are small, whereas times~$s$ and $t$ are large.
The critical regime in higher dimension will be analyzed
in sections~\ref{tcl} and~\ref{tfl} for $D>4$ and $2<D<4$ respectively.
Finally, the dynamics in the low-temperature ferromagnetic phase ($T<T_c$)
will be the subject of section~\ref{ferro}.

\subsection{One-time observables}
\label{t1one}

Let us start the analysis with the functions $f(t)$ and $g(t)$.
In the scaling regime,~(\ref{fglap}) and~(\ref{fsg}) (or~(\ref{flap1})) yield
\beq
f^\L(p)\approx\frac{1}{\sqrt{8p}},\qquad
g^\L(p)\approx\frac{1}{\sqrt{8p}-4\mu},
\eeq
hence
\beq
f(t)\approx\frac{1}{\sqrt{8\pi t}},\qquad
g(t)\approx\frac{1}{\sqrt{8\pi t}}
+\frac{\mu}{2}\,\e^{2\mu^2t}\bigl(1+\erf(\mu\sqrt{2t})\bigr).
\label{fg1}
\eeq

A quantity of interest in this regime is the reduced dynamical susceptibility
\beq
\chi(t)=\sum_{x=-\infty}^{+\infty}C_x(t)
=C^\F(0,t)=\frac{1}{g(t)}\left(1+4\mu\int_0^t\d s\,g(s)\right).
\label{chit}
\eeq
This quantity provides a measure of the range of the correlations,
i.e., of the typical size of the ordered domains at time $t$.
Using~(\ref{fg1}) and working out the integral in~(\ref{chit}), (or
alternatively using~(\ref{cf1}) below)
we can recast the dynamical susceptibility into the scaling form
\beq
\chi(t)\approx\chi_\st\,F(x),
\label{chisca}
\eeq
where
$\chi_\st\approx{2}/{\mu}$
is the static susceptibility,
and $F(x)$ is the scaling function
\beq
F(x)=\frac{\e^{x^2}(1+\erf x)}{\frac{1}{x\sqrt{\pi}}+\e^{x^2}(1+\erf x)},
\eeq
with $x=\mu\sqrt{2t}=\sqrt{t/\tau_\eq}$.
The scaling law~(\ref{chisca}) interpolates between the coarsening regime
($t\ll\tau_\eq$)
and the stationary regime ($t\gg\tau_\eq$).
In the first case, i.e., for $x\ll1$,
the square-root behavior $F(x)\approx x\sqrt\pi$ yields the coarsening law
\beq
\chi(t)\approx\sqrt{8\pi t}.
\eeq
In the second case, i.e., for $x\gg1$,
the behavior $F(x)\approx 1-2\e^{-x^2}/(x\sqrt{\pi})$
yields an exponential convergence of the dynamical susceptibility
to its static value $\chi_\st$.

To close, let us mention that the full structure factor $C^\F(q,t)$
can be derived explicitly in the scaling regime.
Starting from~(\ref{cq}) and using again~(\ref{fg1}), we get
\beqa
C^\F(q,t)&\approx&\frac{\e^{-2q^2t}}{g(t)}\left(1+4\mu\int_0^t\d s
\,\e^{2q^2s}g(s)\right)
\nonumber\\
&\approx&\frac
{\mu^2\,\e^{2\mu^2t}\bigl(1\!+\!\erf(\mu\sqrt{2t})\bigr)
+q\,\e^{-2q^2t}\bigl(q-\ii\mu\erf(\ii q\sqrt{2t})\bigr)}
{(q^2+\mu^2)g(t)}.
\label{cf1}
\eeqa
This expression interpolates continuously between the Gaussian line-shape
\beq
C^\F(q,t)\approx\sqrt{8\pi t}\,\e^{-2q^2t}
\eeq
in the coarsening regime and the Lorentzian one
\beq
C_\eq^\F(q)\approx\frac{2\mu}{q^2+\mu^2}
\eeq
in the stationary state.

\subsection{Two-time observables}
\label{t1two}

Let us now analyze the two-time correlation and response functions
$C(t,s)$ and $R(t,s)$ and the corresponding \fd ratio $X(t,s)$ in the scaling regime.
Starting from~(\ref{cq}),~(\ref{cqts}) and~(\ref{rqts}),
and performing the Gaussian integrals over~$q$, we obtain
\beqa
C(t,s)&\approx&\frac{1}{\sqrt{g(t)g(s)}}
\nonumber\\
&\times&\left(\frac{\e^{-V^2(t-s)^2/(t+s)}}{\sqrt{4\pi(t+s)}}
+4\mu\int_0^s\d u\,\frac{\e^{-V^2(t-s)^2/(t+s-2u)}}{\sqrt{4\pi(t+s-2u)}}g(u)\right),
\nonumber\\
R(t,s)&\approx&\frac{\e^{-V^2(t-s)}}{\sqrt{4\pi(t-s)}}\sqrt{\frac{g(s)}{g(t)}},
\label{cr1gene}
\eeqa
where the function $g(t)$ is given in~(\ref{fg1}).
Many different sub-cases can be analyzed from the general
expressions~(\ref{cr1gene}),
according to the values of the dimensionless combinations $\mu^2s$,
$\mu^2t$, $V^2s$ and $V^2t$.
The most interesting of these regimes are as follows.

\subsubsection*{\it Stationary regime.}
This regime is reached when the waiting time $s$ is much larger than $\tau_\eq$
($\mu^2s\gg1$).
The function $g(t)$ obeys the exponential growth~(\ref{ghigh}).
The corresponding two-time quantities have been studied in
section~\ref{stat1low}.

\subsubsection*{Coarsening regime.}
This regime corresponds to the zero-temperature limit
of the reversible dynamics ($\mu=V=0$).
The resulting scale-invariant coarsening dynamics has been investigated
earlier~\cite{gl2D}.
The expressions~(\ref{cr1gene}) simplify~to
\beqa
C(t,s)&\approx&\left(\frac{4ts}{(t+s)^2}\right)^{1/4},
\nonumber\\
R(t,s)&\approx&\frac{1}{\sqrt{4\pi(t-s)}}\left(\frac{t}{s}\right)^{1/4}.
\label{cr1zero}
\eeqa
The corresponding \fd ratio reads
\beq
X(t,s)\approx
T\left(\frac{2s}{\pi}\right)^{1/2}\left(\frac{t+s}{t-s}\right)^{3/2},
\eeq
and especially (see~(\ref{xasdef}))
\beq
X_\as(s)\approx T\left(\frac{2s}{\pi}\right)^{1/2}.
\label{xcoa}
\eeq

\subsubsection*{Anomalous aging.}
This regime corresponds to the zero-temperature limit
of the irreversible dynamics ($\mu=0$, while $V\ne0$).
The expressions~(\ref{cr1gene}) simplify~to
\beqa
C(t,s)&\approx&\left(\frac{4ts}{(t+s)^2}\right)^{1/4}
\e^{-V^2(t-s)^2/(t+s)},
\nonumber\\
R(t,s)&\approx&\frac{\e^{-V^2(t-s)}}{\sqrt{4\pi(t-s)}}\left(\frac{t}{s}\right)^{1/4}.
\label{cr1anom}
\eeqa
The response function $R(t,s)$ exhibits an exponential decay in $\tau=t-s$,
characterized by the microscopic time scale $1/V^2$.
The correlation function $C(t,s)$ however exhibits a richer behavior
in the late-time regime ($V^2s\gg1$),
with an intermediate {\it anomalous aging regime}~\cite{lm} for $\tau\ll s$,
where it assumes the Gaussian form
\beq
C(t,s)\approx\e^{-V^2\tau^2/(2s)},
\label{ano1}
\eeq
which falls over a characteristic time growing as $\sqrt{s}/V$.
When $\tau$ becomes of the order of the waiting time $s$,
the correlation function is already exponentially small in the variable $V^2s$.
Its subsequent decay to zero follows an exponential law,
characterized by the same microscopic time scale $1/V^2$ as the response function.
The same regime can be observed in the case of the Ising chain with asymmetric
dynamics~\cite{cg2011}.

\subsubsection*{Asymptotic regime.}
This regime corresponds to the limit of a very long time difference $\tau=t-s$,
while the combinations $\mu^2s$ and $V^2s$ remain finite.
It is the non-stationary counterpart
of the large-$\tau$ stationary regime investigated in section~\ref{stat1}.
Here again,
rather than using the expressions~(\ref{cr1gene}),
it is advantageous to return to the integral expressions~(\ref{cts})
and~(\ref{rts}).
For large $\tau$,
the latter integrals are dominated by a saddle point at $q\un=-\ii V$.
We thus obtain the following estimates:
\beqa
C(t,s)&\approx&\frac{\e^{-(\mu^2+V^2)\tau}}{\sqrt{4\pi\tau}}\,
\e^{-\mu^2s}\sqrt\frac{g(s)}{\mu}\,C^\F(-\ii V,s),
\nonumber\\
R(t,s)&\approx&\frac{\e^{-(\mu^2+V^2)\tau}}{\sqrt{4\pi\tau}}\,
\e^{-\mu^2s}\sqrt\frac{g(s)}{\mu}.
\label{cras}
\eeqa
The structure factor $C^\F(-\ii V,s)$ entering the first of these expressions
is given by the analytic continuation of~(\ref{cf1}).

One of the most physically relevant quantities
characterizing the transient regime, before the stationary state sets in,
is the asymptotic \fd ratio $X_\as(s)$ (see~(\ref{xasdef})).
Defining the scaling variables $x$ and $y$ by
\beq
x=\mu\sqrt{2s}$,\qquad $y=V\!\sqrt{2s},
\label{xy1}
\eeq
the expressions~(\ref{cras}), together with~(\ref{fg1}) and~(\ref{cf1}),
yield the scaling formula
\beq
\frac{1}{X_\as(s)}\approx2+\frac{N(x,y)}
{2x(x^2-y^2)\left(\frac{1}{\sqrt{\pi}}+x\,\e^{x^2}(1+\erf x)\right)^2},
\label{xas1}
\eeq
with
\beqa
N(x,y)&=&\left(\frac{1+2(3y^2-x^2)}{\sqrt{\pi}}+2x(3y^2-x^2)\,\e^{x^2}(1+\erf x)\right)
\nonumber\\
&\times&\left(x^2\,\e^{x^2}(1+\erf x)-y\,\e^{y^2}(y+x\erf y)\right).
\eeqa
In the borderline situation where $x=y$, i.e., $V=V_\B=\mu$,
the above expression simplifies to
\beq
\frac{1}{X_\as(s)}\approx2+\frac{N_0(x)}
{4x\left(\frac{1}{\sqrt{\pi}}+x\,\e^{x^2}(1+\erf x)\right)^2},
\label{xasb}
\eeq
with
\beqa
N_0(x)&=&\left(\frac{1+4x^2}{\sqrt{\pi}}+4x^3\,\e^{x^2}(1+\erf x)\right)
\nonumber\\
&\times&\left(\frac{2x}{\sqrt{\pi}}+2(x^2+1)\,\e^{x^2}+(2x^2+1)\,\e^{x^2}\erf x\right).
\eeqa

At the coarsening end,
i.e., when $x$ and $y$ are small, the \fd ratio has the series expansion
\beqa
X_\as(s)&\approx&\frac{2}{\sqrt{\pi}}\,x+\frac{4(\pi-3)}{\pi}\,x^2
+\frac{2(\pi-3)(\pi-12)}{\pi^{3/2}}\,x^3
\nonumber\\
&-&\frac{14}{\sqrt{\pi}}\,xy^2+\cdots
\label{xcoafl}
\eeqa
The first term agrees with~(\ref{xcoa}),
whereas only the last one shows a dependence on~$V$.

At the stationary end,
i.e., when $x$ and $y$ are large,
the distinction between the stationary Regimes~I and II shows up as follows.

\begin{itemize}

\item
For $y<x$, i.e., $V<\mu$, we have
\beq
X_\as(s)\approx\frac{x^2-y^2}{x^2+y^2}
\left(1+\frac{y(x+y)(3y^2-x^2)}{2x^2(x^2+y^2)}\,\e^{-(x^2-y^2)}\right).
\eeq
The limit value of this expression is in agreement with the
stationary ratio $X_\st(\infty)$ in Regime~I, given by~(\ref{xinfstat}).
In other words, we have
\beq
\lim_{s\to\infty}X_\as(s)=\lim_{\tau\to\infty}X_\st(\tau).
\eeq
This identity testifies that the convergence of the \fd ratio
to the value $X_\st(\infty)$ is uniform in the $s$--$t$ plane,
irrespective of the direction,
provided both times $s$ and $t$ are much larger than $\tau_\eq$ and $1/V^2$.

The \fd ratio $X_\as(s)$ increases monotonically and reaches its limit from below
for $y/x<1/\sqrt3$, i.e., $1/2<X_\as(\infty)<1$,
whereas it exhibits a maximum and reaches its limit from above
for $1/\sqrt3<y/x<1$, i.e., $X_\as(\infty)<1/2$.
This feature will also hold in higher dimension,
where the value $X_\as(\infty)=1/2$ will always be singled out.

\item
For $y>x$, i.e., $V>\mu$, the \fd ratio falls off exponentially as
\beq
X_\as(s)\approx\frac{2x^2(y-x)}{y(x+y)(3y^2-x^2)}\,\e^{-(y^2-x^2)}.
\eeq
This corresponds to the stationary Regime~II.

\item
In the borderline situation where $x=y$, i.e., $V=V_\B=\mu$,
the \fd ratio, given by~(\ref{xasb}), falls off as a power law:
\beq
X_\as(s)\approx\frac{1}{2x^2}\approx\frac{1}{4\mu^2s}.
\eeq

\end{itemize}

\begin{figure}[!ht]
\begin{center}
\includegraphics[angle=0,width=.7\linewidth]{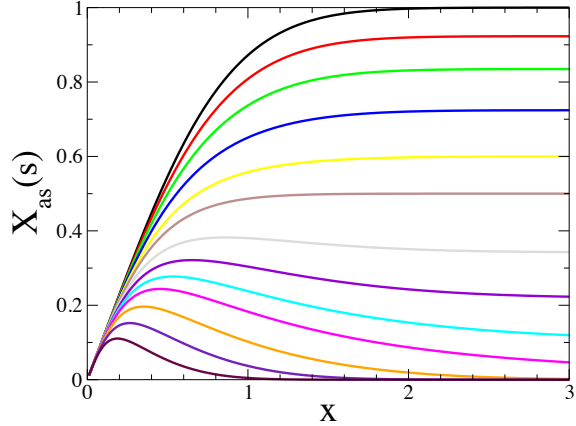}
\caption{\small
Plot of the asymptotic \fd ratio $X_\as(s)$
in the low-temperature scaling regime in one dimension (see~(\ref{xas1}))
against the scaling variable $x=\mu\sqrt{2s}$,
for the following values of the ratio $V/\mu$ (top to bottom):
0 (reversible), 0.2, 0.3, 0.4, 0.5,
$1/\sqrt3\approx0.577350$ (so that $X_\st(\infty)=1/2$),
0.7, 0.8, 0.9, 1 (borderline case), 1.2, 1.5 and 2.}
\label{x1}
\end{center}
\end{figure}

Figure~\ref{x1} shows a plot of the asymptotic \fd ratio $X_\as(s)$
against the scaling variable $x=\mu\sqrt{2s}$,
for several values of the ratio $y/x=V/\mu$.

\section{Classical critical regime ($D>4$)}
\label{tcl}

In this section we study the transient properties of the model
in the vicinity of its critical temperature for $D>4$,
where the critical exponent $\nu=1/2$ assumes its mean-field or `classical' value.
We assume that $\mu$ and $\V$ are small, whereas times~$s$ and $t$ are large.
The dynamical properties of the model obey scaling laws
throughout the scaling regime thus defined.

\subsection{One-time observables}

The constants $A_1$ and $A_2$ defined in~(\ref{adef}) are finite.
As a consequence,~(\ref{fglap}) and~(\ref{freg}) yield in the scaling regime
\beq
g^\L(p)\approx\frac{G}{p-2\mu^2},\qquad
G=\frac{A_1^2}{A_2},
\eeq
hence
\beq
g(t)\approx G\,\e^{2\mu^2t}.
\label{gcl}
\eeq
The structure factor $C^\F(\q,t)$ can be derived explicitly in the scaling regime.
Keeping only the integral term in~(\ref{cq}) and using~(\ref{gcl}), we get
\beq
C^\F(\q,t)\approx T_c\,\frac{1-\e^{-2(\q^2+\mu^2)t}}{\q^2+\mu^2}.
\label{cfcl}
\eeq
The reduced dynamical susceptibility $\chi(t)=C^\F(\0,t)$ therefore reads
\beq
\chi(t)\approx T_c\,\frac{1-\e^{-2\mu^2t}}{\mu^2}.
\eeq
This quantity exhibits a linear growth $\chi(t)\approx2T_ct$
in the coarsening regime,
and saturates to the static value $\chi_\st=T_c/\mu^2$
in the stationary state.

\subsection{Two-time observables}

In order to study the two-time correlation and response functions,
we again start from~(\ref{cq}),~(\ref{cqts}) and~(\ref{rqts}),
use~(\ref{gcl}), and perform the Gaussian integrals over~$\q$.
We thus obtain
\beqa
C(t,s)&\approx&2T_c\,\e^{-\mu^2(t-s)}
\int_0^s\d u\,\frac{\e^{-2\mu^2(s-u)-\V^2(t-s)^2/(t+s-2u)}}{(4\pi(t+s-2u))^{D/2}},
\nonumber\\
R(t,s)&\approx&\frac{\e^{-(\mu^2+\V^2)(t-s)}}{(4\pi(t-s))^{D/2}}.
\label{crcl}
\eeqa
The response function does not exhibit any aging property:
it coincides with the continuum limit of its stationary expression~(\ref{rhigh}).
Many different sub-cases follow from the general expressions~(\ref{crcl}).
We again restrict the study to the most interesting ones.

\subsubsection*{Critical coarsening.}
This scale-invariant regime corresponds to the reversible dyn\-amics at the critical
point ($\mu=0$, $\V=\0$)~\cite{gl2D}.
The expressions~(\ref{crcl}) simplify to
\beqa
C(t,s)&\approx&\frac{2T_c}{(D-2)(4\pi)^{D/2}}
\left((t-s)^{-(D-2)/2}-(t+s)^{-(D-2)/2}\right),
\nonumber\\
R(t,s)&\approx&\frac{1}{(4\pi(t-s))^{D/2}}.
\eeqa
The corresponding \fd ratio,
\beq
X(t,s)\approx\frac{1}{1+\left(\frad{t-s}{t+s}\right)^{D/2}},
\eeq
only depends on the ratio $t/s$.
In the limit where this time ratio is very large,
the \fd ratio tends to a finite value
\beq
X_\infty=\lim_{s\to\infty}\lim_{\tau\to\infty}X(t,s)=\frac{1}{2},
\label{xinfcl}
\eeq
which has been emphasized~\cite{gl1D,gl2D} to be a new universal amplitude ratio
pertaining to nonequilibrium critical dynamics.

\subsubsection*{Anomalous aging.}
This regime corresponds to the irreversible dynamics
at the critical point ($\mu=0$, while $\V\ne\0$).
As in one dimension, the two-point correlation function exhibits
an intermediate anomalous aging regime for $\tau\ll s$.
The scaling form of $C(t,s)$ is however more complex than
the Gaussian form~(\ref{ano1}).
A scaling analysis of the integral entering~(\ref{crcl}),
neglecting $\tau$ with respect to $s$ and integrating over $z=s/u$,
indeed yields
\beqa
C(t,s)&\approx&\frac{2T_c}{(8\pi)^{D/2}s^{(D-2)/2}}
\int_1^\infty\d z\,\e^{-\V^2\tau^2z/(2s)}z^{(D-4)/2}
\nonumber\\
&\approx&\frac{T_c}{(4\pi)^{D/2}(\abs{\V}\tau)^{D-2}}\;
\Gamma\left(\frac{D-2}{2},\frac{\V^2\tau^2}{2s}\right),
\label{anocl}
\eeqa
where
\beq
\Gamma(a,x)=\int_x^\infty\d y\,\e^{-y}y^{a-1}
\eeq
is the complementary incomplete Gamma function.

\subsubsection*{Asymptotic regime.}
This asymptotic regime is the non-stationary counterpart
of the large-$\tau$ stationary regime of section~\ref{stathigh}.
The two-time response function~(\ref{crcl}) is already in asymptotic form,
whereas for large $\tau$
the integral expression~(\ref{cts}) for the two-time correlation function
is dominated by a saddle point at $\q\un=-\ii\V$.
Using~(\ref{gcl}) and~(\ref{cfcl}), we obtain
\beq
C(t,s)\approx T_c\frac{\e^{-(\mu^2+\V^2)\tau}}{(4\pi\tau)^{D/2}}\,
\frac{1-\e^{-2(\mu^2-\V^2)s}}{\mu^2-\V^2}.
\label{cascl}
\eeq

The asymptotic \fd ratio $X_\as(s)$ is therefore
\beq
X_\as(s)\approx\frac{\mu^2-\V^2}{\mu^2+\V^2+(\mu^2-3\V^2)\,\e^{-2(\mu^2-\V^2)s}}.
\label{xascl}
\eeq
Introducing the the scaling variables
\beq
x=\mu\sqrt{2s},\qquad
y=\abs{\V}\sqrt{2s},
\label{xyd}
\eeq
(see~(\ref{xy1})),
the above result reads
\beq
X_\as(s)\approx\frac{x^2-y^2}{x^2+y^2+(x^2-3y^2)\,\e^{-(x^2-y^2)}}.
\label{xasclsca}
\eeq

At the critical end,
i.e., when $\mu^2s$ and $\V^2s$ are small, the \fd ratio has the series expansion
\beq
X_\as(s)\approx\frac{1}{2}\left(1+(\mu^2-3\V^2)\,s+\cdots\right),
\label{xcoacl}
\eeq
whose first term agrees with~(\ref{xinfcl}).

At the stationary end,
the distinction between the stationary Regimes~I and II shows up as follows.

\begin{itemize}

\item
For $\V^2<\mu^2$, we have
\beq
X_\as(s)\approx\frac{\mu^2-\V^2}{\mu^2+\V^2}
\left(1+\frac{3\V^2-\mu^2}{\mu^2+\V^2}\,\e^{-2(\mu^2-\V^2)s}\right).
\eeq
The limit value is in agreement with the stationary $X_\st(\infty)$
(see~(\ref{xinfscahigh})) in Regime~I.
The \fd ratio increases monotonically and reaches its limit from below
for $\V^2<\mu^2/3$, i.e., $1/2<X_\st(\infty)<1$,
whereas it decreases monotonically and reaches its limit from above
for $\V^2>\mu^2/3$, i.e., $X_\st(\infty)<1/2$.
In the special case where $\V^2=\mu^2/3$,
i.e., $y=x/\sqrt3$, we have $X_\as(s)=1/2$ for all values of $s$.

\item
For $\V^2>\mu^2$, the \fd ratio falls off exponentially as
\beq
X_\as(s)\approx\frac{\V^2-\mu^2}{3\V^2-\mu^2}\,\e^{-2(\V^2-\mu^2)s}.
\eeq
This corresponds to the stationary Regime~II.

\item
In the borderline situation where $\V^2=\mu^2$, the \fd ratio again falls off as
\beq
X_\as(s)\approx\frac{1}{2x^2}\approx\frac{1}{4\mu^2s}.
\eeq

\end{itemize}

\begin{figure}[!ht]
\begin{center}
\includegraphics[angle=0,width=.7\linewidth]{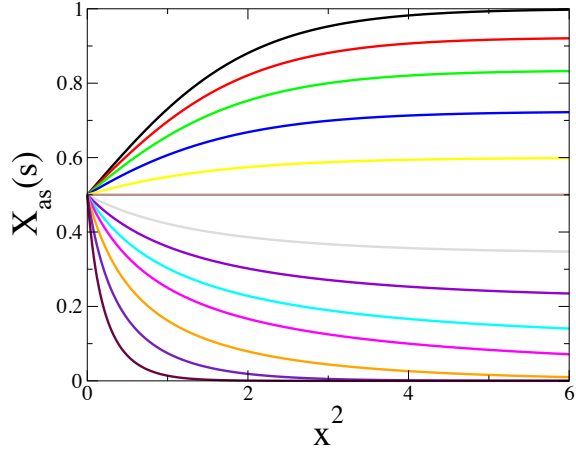}
\caption{\small
Plot of the asymptotic \fd ratio $X_\as(s)$
in the classical critical regime ($D>4$) (see~(\ref{xascl}))
against $x^2=2\mu^2s$,
for the following values of the ratio $\abs{\V}/\mu$ (top to bottom):
0 (reversible), 0.2, 0.3, 0.4, 0.5,
$1/\sqrt3\approx0.577350$ (so that $X_\as(s)=1/2$ for all $s$),
0.7, 0.8, 0.9, 1 (borderline case), 1.2, 1.5 and 2.}
\label{xcl}
\end{center}
\end{figure}

Figure~\ref{xcl} shows a plot of the asymptotic \fd ratio $X_\as(s)$
against $x^2=2\mu^2s$, for several values of the ratio $\abs{\V}/\mu$.

\section{Non-classical critical regime ($2<D<4$)}
\label{tfl}

In this section we study the transient properties of the model
in the vicinity of its critical temperature for $2<D<4$,
where the critical exponent $\nu=1/(D-2)$ depends continuously on dimension.
We again assume that $\mu$ and $\V$ are small, whereas times~$s$ and $t$ are large.

\subsection{One-time observables}
\label{tflone}

Let us, for convenience, parametrize dimension as
\beq
D=2(1+\eps)\qquad(0<\eps<1),
\eeq
so that $\nu=1/(2\eps)$.
In the scaling regime,~(\ref{fglap}) and~(\ref{freg}) yield
\beq
f^\L(p)\approx A_1-Cp^\eps,\qquad
g^\L(p)\approx\frac{G}{p^\eps-(2\mu^2)^\eps},
\eeq
with
\beq
C=(8\pi)^{-1-\eps}\abs{\Gamma(-\eps)},\qquad G=\frac{A_1^2}{C}.
\eeq

We have therefore
\beq
g(t)\approx\frac{G}{\eps}(2\mu^2)^{1-\eps}\,h_\eps(2\mu^2t),
\label{gfl}
\eeq
where the scaling function $h_\eps$ reads\footnote{$\C_1$ (resp.~$\C_0$)
is a vertical contour in the $z$-plane such that $\Re z>1$ (resp.~$0<\Re z<1$).}
\beq
h_\eps(\xi)=\int_{\C_1}\frac{\d z}{2\pi\ii}\,\frac{\eps}{z^\eps-1}\,\e^{z\xi}
=\e^\xi+\int_{\C_0}\frac{\d z}{2\pi\ii}\,\frac{\eps}{z^\eps-1}\,\e^{z\xi}.
\eeq
By folding the contour $\C_0$ around the cut
(i.e., $z\to-\rho\pm\ii 0$), we obtain the integral representation
\beq
h_\eps(\xi)=\e^\xi+\frac{\eps\,\sin\pi\eps}{\pi}
\int_0^\infty\d\rho\,\frac{\e^{-\rho\xi}}{\rho^\eps+\rho^{-\eps}-2\cos\pi\eps}.
\eeq
Finally, the series expansion
\beq
h_\eps(\xi)=\eps\sum_{n\ge1}\frac{\xi^{n\eps-1}}{\Gamma(n\eps)}
=\eps\frac{\d}{\d\xi}E_\eps(\xi^\eps)
\eeq
relates $h_\eps$ to the Mittag-Leffler function $E_\eps$.

In the coarsening regime,
corresponding to small values of the scaling variable $\xi$,
the scaling function $h_\eps$ diverges as a power law:
\beq
h_\eps(\xi)\approx\frac{\eps}{\Gamma(\eps)}\,\xi^{\eps-1}.
\label{hlow}
\eeq
At the stationary end, corresponding to large values of $\xi$,
it grows exponentially as
\beq
h_\eps(\xi)\approx\e^\xi.
\label{hhigh}
\eeq

The behavior of the reduced dynamical susceptibility $\chi(t)=C^\F(\0,t)$
in the scaling regime
can be estimated by keeping only the integral term in~(\ref{cq}) and using~(\ref{gfl}).
We thus obtain
\beq
\chi(t)\approx\frac{2T_c}{h_\eps(2\mu^2t)}\int_0^t\d s\, h_\eps(2\mu^2s).
\eeq
In the coarsening regime (see~(\ref{hlow})),
the susceptibility exhibits the linear growth $\chi(t)\approx(2T_ct)/\eps$.
At the stationary end (see~(\ref{hhigh})),
the susceptibility saturates to its static value $\chi_\st=T_c/\mu^2$.

The following special values of $\eps$ are of interest,
as they correspond to integer dimensions.
We shall return to them at the end of section~\ref{tfltwo}.

\begin{itemize}

\item
$D=2$ (i.e., $\eps=0$).
This case corresponds to the lower critical dimension of the model.
We have
\beq
h_0(\xi)=\int_0^\infty\d x\,\frac{\xi^{x-1}}{\Gamma(x)}
=\e^\xi+\int_0^\infty\d\rho\,\frac{\e^{-\rho\xi}}{(\ln\rho)^2+\pi^2}.
\label{hzero}
\eeq
The equality between the two integral expressions is attributed to
Ramanujan~\cite{hardy}.

\item
$D=3$ (i.e., $\eps=1/2$).
This case is the only non-trivial one with integer dimension.
The scaling function reads
\beq
h_{1/2}(\xi)=\frac{1}{\sqrt{4\pi\xi}}+\frac12\e^\xi(1+\erf\!\sqrt\xi).
\label{hhalf}
\eeq

\item
$D=4$ (i.e., $\eps=1$).
This case corresponds to the upper critical dimension of the model.
The scaling function becomes a simple exponential:
\beq
h_1(\xi)=\e^\xi
\label{hone}
\eeq
(see the comment at the very end of section~\ref{tfltwo}).

\end{itemize}

\subsection{Two-time observables}
\label{tfltwo}

Starting from~(\ref{cq}),~(\ref{cqts}) and~(\ref{rqts})
and performing the Gaussian integrals over~$\q$,
we are left with the following expressions
for the two-time correlation and response functions in the scaling regime:
\beqa
C(t,s)&\approx&\frac{2T_c}{\sqrt{h_\eps(2\mu^2t)h_\eps(2\mu^2s)}}
\int_0^s\d u\,\frac{\e^{-\V^2(t-s)^2/(t+s-2u)}}{(4\pi(t+s-2u))^{D/2}}
\,h_\eps(2\mu^2u),
\nonumber\\
R(t,s)&\approx&\frac{\e^{-\V^2(t-s)}}{(4\pi(t-s))^{D/2}}
\sqrt{\frac{h_\eps(2\mu^2s)}{h_\eps(2\mu^2t)}}.
\label{crfl}
\eeqa
In what follows we restrict the study to the
most interesting of the different sub-cases which can be investigated from
these general expressions,
according to the values of the dimensionless combinations $\mu^2s$,
$\mu^2t$, $\V^2s$ and $\V^2t$.

\subsubsection*{Critical coarsening.}
This scale-invariant regime corresponds to the reversible dyn\-amics at the critical
point ($\mu=0$, $\V=\0$)~\cite{gl2D}.
Owing to~(\ref{hlow}), the expressions~(\ref{crfl}) simplify~to
\beqa
C(t,s)&\approx&\frac{4T_c}{(D-2)(4\pi)^{D/2}(t-s)^{(D-2)/2}}
\frac{t}{t+s}\left(\frac{t}{s}\right)^{-D/4},
\nonumber\\
R(t,s)&\approx&\frac{1}{(4\pi(t-s))^{D/2}}\left(\frac{t}{s}\right)^{1-D/4}.
\eeqa
The corresponding \fd ratio reads
\beq
X(t,s)\approx\frac{1}{1+\frad{2}{D-2}\left(\frad{t-s}{t+s}\right)^2}.
\eeq
The asymptotic \fd ratio assumes the universal value~\cite{gl2D}
\beq
X_\infty=\frac{D-2}{D}.
\label{xinffl}
\eeq

\subsubsection*{Anomalous aging.}
This regime corresponds to the irreversible dynamics
at the critical point ($\mu=0$, while $\V\ne\0$).
The two-point correlation function again exhibits
an intermediate anomalous aging regime for $\tau\ll s$.
A scaling analysis of the integral entering~(\ref{crfl}),
neglecting $\tau$ with respect to $s$ and integrating over the dimensionless
variable $z=s/u$, yields
\beq
C(t,s)\approx\frac{T_c\,\Gamma((D-2)/2)}{(4\pi)^{D/2}(\abs{\V}\tau)^{D-2}}
\,\e^{-\V^2\tau^2/(2s)}.
\label{anofl}
\eeq

\subsubsection*{Asymptotic regime.}
The asymptotic regime is the non-stationary counterpart
of the large-$\tau$ stationary regime analyzed in section~\ref{stathigh}.
For large $\tau$, the integral expressions~(\ref{cts}) and~(\ref{rts})
are dominated by a saddle point at $\q\un=-\ii\V$.
We thus obtain the following estimates:
\beqa
C(t,s)&\approx&\frac{\e^{-(\mu^2+\V^2)\tau}}{(4\pi\tau)^{D/2}}\,
\e^{-\mu^2s}\sqrt{h_\eps(2\mu^2s)}\,C^\F(-\ii\V,s),
\nonumber\\
R(t,s)&\approx&\frac{\e^{-(\mu^2+\V^2)\tau}}{(4\pi\tau)^{D/2}}\,
\e^{-\mu^2s}\sqrt{h_\eps(2\mu^2s)},
\label{crasfl}
\eeqa
with
\beq
C^\F(-\ii\V,s)\approx\frac{2T_c\,\e^{2\V^2s}}{h_\eps(2\mu^2s)}
\int_0^s\d u\,\e^{-2\V^2u}h_\eps(2\mu^2u).
\eeq

The asymptotic \fd ratio $X_\as(s)$ thus reads
\beqa
\frac{1}{X_\as(s)}\approx2&+&\frac{2(3\V^2h_\eps(2\mu^2s)-\mu^2h'_\eps(2\mu^2s))}
{h_\eps(2\mu^2s)^2}
\nonumber\\
&\times&\int_0^s\d u\,\e^{2\V^2(s-u)}h_\eps(2\mu^2u).
\eeqa
Setting
\beq
\xi=x^2=2\mu^2s=\frac{s}{\tau_\eq},\qquad\sigma=\frac{\V^2}{\mu^2}=\frac{y^2}{x^2}
\eeq
(see~(\ref{xyd})),
we obtain the scaling form
\beq
\frac{1}{X_\as(s)}\approx2+\frac{3\sigma h_\eps(\xi)-h'_\eps(\xi)}{h_\eps(\xi)^2}
\int_0^\xi\d\eta\, \e^{\sigma(\xi-\eta)}h_\eps(\eta).
\label{xasfl}
\eeq

At the critical end, i.e., when $x$ and $y$ are small, the \fd ratio has the expansion
\beq
X_\as(s)\approx\frac{\eps}{1+\eps}
\left(1+\frac{\Gamma(\eps)}{2\Gamma(2\eps)}\,x^{2\eps}+\cdots
-\frac{2(2+\eps)}{(1+\eps)^2}\,y^2+\cdots\right).
\eeq
The first term coincides with~(\ref{xinffl}),
whereas the last one exhibits the leading dependence of the result on~$\V$.
For $\eps=1$, i.e., $D=4$, the full expansion agrees with~(\ref{xcoacl}).

At the stationary end,
the distinction between the stationary Regimes~I and II again shows up as follows.

\begin{itemize}

\item
For $\V^2<\mu^2$,
the \fd ratio converges to its stationary value
$X_\st(\infty)$ (see~(\ref{xinfscahigh})).

\item
For $\V^2>\mu^2$, it falls off exponentially as
\beq
X_\as(s)\sim\e^{-2(\V^2-\mu^2)s}.
\eeq

\item
In the borderline situation where $\V^2=\mu^2$,
it again falls off as
\beq
X_\as(s)\approx\frac{1}{4\mu^2s}.
\eeq

\end{itemize}

To close, we analyze in some more detail the asymptotic regime
in the following cases, where the dimension is an integer.

\subsubsection*{The two-dimensional case ($\eps=0$).}
This marginal situation corresponds to the lower critical dimension of the model.
We thus have $T_c=0$,
while $\mu$ vanishes exponentially fast at low temperature, according to~(\ref{mutwo}).
The critical end is affected by the presence of logarithmic corrections
to scaling.
Let us introduce for convenience the logarithmic variable
\beq
\Lambda=\ln\frac{1}{\xi}=\ln\frac{1}{2\mu^2s}=\ln\frac{\tau_\eq}{s},
\eeq
which is large and positive in the coarsening regime $(s\ll\tau_\eq)$.
In this regime, the rightmost expression for $h_0(\xi)$
in~(\ref{hzero}) is dominated by the integral.
Setting $u=\rho\xi$, we obtain the estimate
\beq
\xi\,h_0(\xi)\approx\int_0^\infty\d u\frac{\e^{-u}}{(\Lambda+\ln u)^2+\pi^2},
\eeq
which admits the asymptotic expansion
\beq
\xi\,h_0(\xi)\approx\frac{1}{\Lambda^2}+\frac{2\euler}{\Lambda^3}
+\frac{3\euler^2-\frac{\pi^2}{2}}{\Lambda^4}+\cdots
\eeq
as $\Lambda\to\infty$ (i.e., $\xi\to0$),
where $\euler\approx0.577215$ is Euler's constant.
We are thus left after some algebra with the logarithmic expansion
\beq
X_\as(s)\approx\frac{1}{\Lambda}+\frac{\euler}{\Lambda^2}+\cdots
-\V^2s\left(\frac{8}{\Lambda}+\frac{2(4\euler-3)}{\Lambda^2}+\cdots\right).
\eeq
The asymptotic \fd ratio thus starts increasing as an inverse logarithm of time
in the coarsening regime in dimension two,
whereas its behavior at the stationary end follows the generic pattern,
with its two regimes.

\subsubsection*{The three-dimensional case ($\eps=1/2$).}
More explicit results can be derived in this case,
which is the only non-trivial one with integer dimension.
Inserting the expression~(\ref{hhalf}) of the scaling function $h_{1/2}(\xi)$
into~(\ref{xasfl}),
we obtain the following scaling form for the \fd ratio:
\beq
\frac{1}{X_\as(s)}\approx2+\frac{N(x,y)}
{2(x^2-y^2)\left(\frac{1}{\sqrt{\pi}}+x\,\e^{x^2}(1+\erf x)\right)^2},
\label{xas3}
\eeq
with
\beqa
N(x,y)&=&\left(\frac{1+2(3y^2-x^2)}{\sqrt{\pi}}+2x(3y^2-x^2)\,\e^{x^2}(1+\erf x)\right)
\nonumber\\
&\times&\left(x\,\e^{x^2}(1+\erf x)-\e^{y^2}(x+y\erf y)\right),
\eeqa
where the scaling variables $x$ and $y$ have been defined in~(\ref{xyd}).
In the borderline situation where $x=y$,
the above expression simplifies to
\beq
\frac{1}{X_\as(s)}\approx2+\frac{N_0(x)}
{4x\left(\frac{1}{\sqrt{\pi}}+x\,\e^{x^2}(1+\erf x)\right)^2},
\label{xas3b}
\eeq
with
\beqa
N_0(x)&=&\left(\frac{1+4x^2}{\sqrt{\pi}}+4x^3\,\e^{x^2}(1+\erf x)\right)
\nonumber\\
&\times&\left(\frac{2x}{\sqrt{\pi}}+2x^2\,\e^{x^2}+(2x^2+1)\,\e^{x^2}\erf x\right).
\eeqa
The above expressions bear a strikingly close resemblance
with the results~(\ref{xas1}),~(\ref{xasb})
describing the low-temperature scaling regime in one dimension.

At the critical end, i.e., when $x$ and $y$ are small,
the \fd ratio has the series expansion
\beq
X_\as(s)\approx\frac{1}{3}+\frac{\sqrt{\pi}}{6}\,x+\frac{64-15\pi}{108}\,x^2
-\frac{20}{27}\,y^2+\cdots
\eeq
The first term agrees with~(\ref{xinffl}), i.e., $X_\infty=1/3$ for $D=3$.

\begin{figure}[!ht]
\begin{center}
\includegraphics[angle=0,width=.7\linewidth]{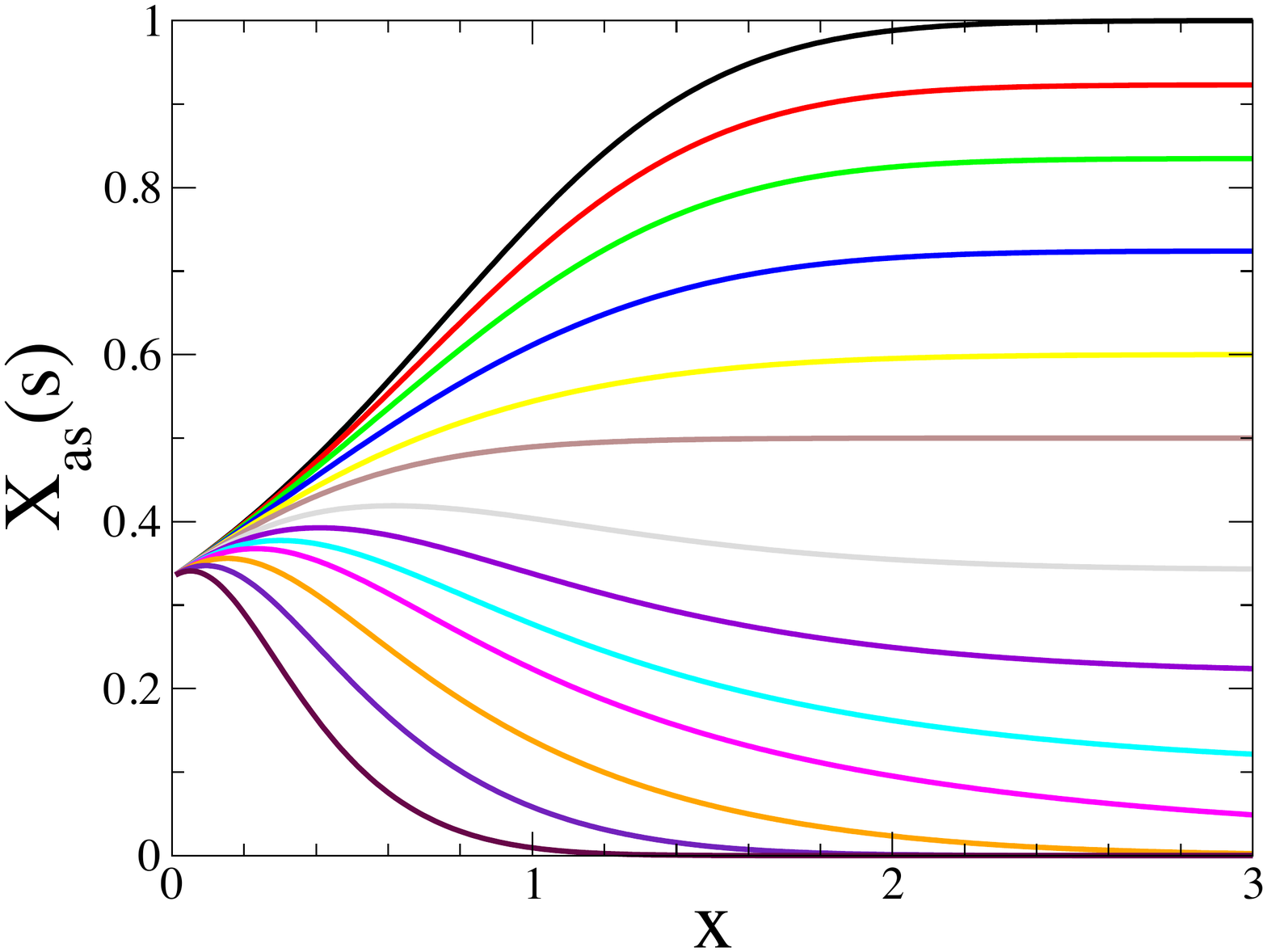}
\caption{\small
Plot of the asymptotic \fd ratio $X_\as(s)$
in the critical regime in dimension three (see~(\ref{xas3}))
against the scaling variable $x=\mu\sqrt{2s}$,
for the following values of the ratio $y/x=\abs\V/\mu$ (top to bottom):
0 (reversible), 0.2, 0.3, 0.4, 0.5,
$1/\sqrt3\approx0.577350$ (so that $X_\st(\infty)=1/2$),
0.7, 0.8, 0.9, 1 (borderline case), 1.2, 1.5 and 2.}
\label{x3}
\end{center}
\end{figure}

Figure~\ref{x3} shows a plot of the asymptotic \fd ratio $X_\as(s)$
against the scaling variable $x=\mu\sqrt{2s}$,
for several values of the ratio $y/x=\abs\V/\mu$.

\subsubsection*{The four-dimensional case ($\eps=1$).}
This situation corresponds to the upper critical dimension of the model,
where the exponent $\nu$ changes from non-classical to classical
(see~(\ref{nures})),
so that logarithmic corrections to scaling can be expected a priori.
The scaling function $h_1(\xi)$ is, however, a simple exponential (see~(\ref{hone})).
As a consequence, to leading order in the critical regime,
the quantities which can be expressed in terms of $h_1(\xi)$
are given for $D=4$ by their expressions in the classical regime ($D>4$),
derived in section~\ref{tcl}.
In other words, logarithmic corrections to scaling are subleading for these quantities.
This phenomenon was already put forward by Ebbinghaus et al~\cite{4dscaling}.
In particular, the asymptotic \fd ratio is given by~(\ref{xascl}),
up to correction terms which are of the order of $1/\ln(1/\mu^2)$,
i.e., $1/\abs{\ln(T-T_c)}$, in relative value.

\section{Ferromagnetic phase}
\label{ferro}

This last section is devoted to the two-time observables of the model
in its low-temperature ferromagnetic phase ($T<T_c$).
This ordered phase only exists in high enough dimension ($D>2$).
It is characterized by a non-zero value of the
spontaneous magnetization~$M_\eq$, given by~(\ref{mdef}).

The dynamical properties of the model in the late-time coarsening regime
are ruled by the power-law decay~(\ref{glow}) of the function $g(t)$.
In particular,
the behavior of the reduced dynamical susceptibility $\chi(t)=C^\F(\0,t)$
can be estimated by keeping only the integral term in~(\ref{cq}),
and using~(\ref{glow}) and the sum rule~(\ref{fgid}).
We thus obtain
\beq
\chi(t)\approx M_\eq^2 (8\pi t)^{D/2}.
\eeq
This result has a simple interpretation:
the typical linear size of a magnetized domain grows
proportionally to the diffusive scale $L(t)\sim\sqrt{t}$~\cite{bray},
so that its volume, measured by $\chi(t)$, grows as $L(t)^D\sim t^{D/2}$.

The commonly accepted phenomenology of glassy dynamics
(see~\cite{glass} for reviews) tells us that
two-time quantities behave differently in the following two regimes,
which will be studied successively hereafter:

\begin{itemize}

\item
a {\it beta (stationary) regime} ($\tau=t-s\ll s$),
where the two-time correlation function $C(t,s)$ only depends on $\tau$,
and decays from unity to its plateau value~$M_\eq^2$.

\item
an {\it alpha (aging) regime} ($\tau\sim s$),
where the two-time correlation function
further decays from its plateau value $M_\eq^2$ to zero.

\end{itemize}

\subsection{Beta (stationary) regime}

In this first regime, i.e., for $\tau\ll s$,
it is legitimate to simplify~(\ref{cqts}) and~(\ref{rqts}) to
\beqa
C_\bbeta^\F(\q,\tau)&\approx&\frac{T}{\omega(\q)}\,\e^{-\Omega(\q)\tau},
\nonumber\\
R_\bbeta^\F(\q,\tau)&\approx&\e^{-\Omega(\q)\tau}.
\label{crbeta}
\eeqa
These expressions coincide with the $\mu\to0$ limit of~(\ref{crst}),
which was shown to also describe the nonequilibrium stationary state at the critical point.

The correlation function therefore behaves as
\beq
C_\bbeta(\tau)\approx M_\eq^2+(1-M_\eq^2)\Phi(\tau),
\eeq
where the function
\beq
\Phi(\tau)=T_c\int\dq\frac{\e^{-\Omega(\q)\tau}}{\omega(\q)}
\eeq
decreases from $\Phi(0)=1$ to $\Phi(\infty)=0$,
and thus describes the relaxation of correlations throughout the beta regime.

Hereafter we focus our attention onto the scaling regime where $\V$ is small.
The response function assumes the simple expression
\beq
R_\bbeta(\tau)\approx\frac{\e^{-\V^2\tau}}{(4\pi\tau)^{D/2}},
\eeq
and therefore falls off exponentially to zero.

More interestingly, the tail of the function $\Phi(\tau)$,
describing how the correlation function tends to its plateau value $M_\eq^2$,
assumes the scaling form
\beq
\Phi(\tau)\approx\frac{T_c}{(4\pi)^{D/2}\tau^{(D-2)/2}}\,F(z),
\eeq
where the scaling variable is
\beq
z=\V^2\tau,
\eeq
and with
\beq
F(z)=z^{-(D-2)/2}\;\gamma\left(\frac{D-2}{2},z\right),
\eeq
where
\beq
\gamma(a,x)=\int_0^x\d y\,\e^{-y}y^{a-1}
\eeq
is the incomplete Gamma function.
The scaling function $F(z)$ starts from the finite value $F(0)=2/(D-2)$,
and decays as $F(z)\approx\Gamma((D-2)/2)z^{-(D-2)/2}$ at large~$z$.
As a consequence, we have
\beq
\Phi(\tau)\approx\frac{2T_c}{(D-2)(4\pi)^{D/2}\tau^{(D-2)/2}}
\label{phis}
\eeq
for reversible dynamics ($\V=\0$),
and
\beq
\Phi(\tau)\approx\frac{T_c\,\Gamma((D-2)/2)}{(4\pi)^{D/2}(\abs{\V}\tau)^{D-2}}
\label{phias}
\eeq
as soon as $\V\ne\0$.

The correlation function therefore always exhibits a slow
power-law convergence toward its plateau value.
The exponent of this power law however depends on whether
the dynamics is reversible or not.
In the first case (see~(\ref{phis})), it equals $(D-2)/2$;
in the second case (see~(\ref{phias})), its value is doubled to $D-2$.

Finally, in the same scaling regime where $\V$ is small and $\tau$ is large,
the \fd ratio also assumes a scaling form, namely
\beq
X_\bbeta(\tau)\approx\X(z),
\eeq
with
\beq
\X(z)=\frac{2\,\e^{-z}}{(D-2)F(z)-2zF'(z)}.
\eeq
The \fd ratio is identically equal to unity for reversible dynamics
($\V=\0$) throughout the beta regime.
In the irreversible case, it starts decreasing as
\beq
\X(z)=1-\frac{4z}{D}+\cdots,
\eeq
i.e., $X_\bbeta(\tau)\approx1-4\V^2\tau/D$,
whereas it falls off roughly exponentially at large $z$.

\begin{figure}[!ht]
\begin{center}
\includegraphics[angle=0,width=.7\linewidth]{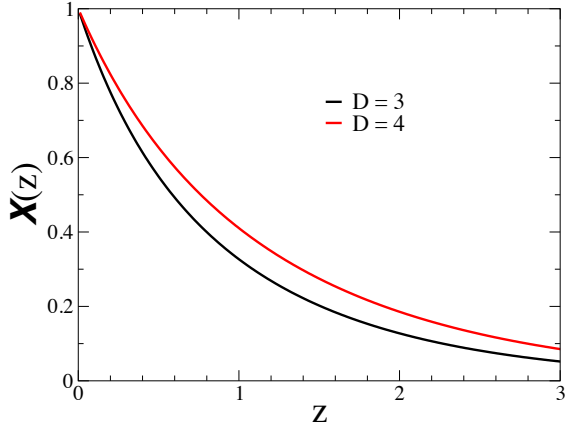}
\caption{\small
Plot of the scaling function $\X(z)$ describing the \fd ratio
in the beta regime in the ferromagnetic phase,
against $z=\V^2\tau$, for $D=3$ and $D=4$.}
\label{x34}
\end{center}
\end{figure}

To close, we give explicit expressions for the scaling functions $F(z)$ and $\X(z)$
in dimension three and four:
\beqa
D=3:\quad&&F(z)=\sqrt\frac{\pi}{z}\erf\sqrt{z},
\quad\X(z)=\frad{\sqrt{z}}{\sqrt{\pi}\e^z\erf\sqrt{z}-\sqrt{z}},
\nonumber\\
D=4:\quad&&F(z)=\frac{1-\e^{-z}}{z},
{\hskip 23pt}\X(z)=\frac{z}{2(\e^z-1)-z}.
\label{fx}
\eeqa
Figure~\ref{x34} shows a plot of the scaling function $\X(z)$ for $D=3$ and $D=4$.

\subsection{Alpha (aging) regime}

Let us again consider, for the sake of simplicity,
the scaling regime where $\V$ is small.
Throughout the alpha (aging) regime,
where both times $s$ and $t$ are large and comparable,
(\ref{cq}),~(\ref{cqts}) and~(\ref{rqts}),
together with~(\ref{glow}) and the sum rule~(\ref{fgid}),
lead to the estimates
\beqa
C_\aalpha(t,s)&\approx&M_\eq^2\left(\frac{4ts}{(t+s)^2}\right)^{D/4}
\e^{-\V^2(t-s)^2/(t+s)},
\nonumber\\
R_\aalpha(t,s)&\approx&\frac{\e^{-\V^2(t-s)}}{(4\pi(t-s))^{D/2}}\left(\frac{t}{s}\right)^{D/4},
\label{crage}
\eeqa
which bear a strong resemblance with the formulas~(\ref{cr1anom})
pertaining to the zero-temperature dynamics of the one-dimensional model.

The corresponding \fd ratio reads
\beqa
X_\aalpha(t,s)&\approx&\frac{4T}{D(8\pi)^{D/2}M_\eq^2}
\left(\frac{t+s}{t-s}\right)^{(D+2)/2}s^{-(D-2)/2}
\nonumber\\
&\times&\frac{D(t+s)}{D(t+s)+4\V^2s(3t+s)}\,\e^{-2\V^2s(t-s)/(t+s)}.
\eeqa
The first line is the known result for reversible dynamics~\cite{gl2D,horner},
whereas the second one describes the effect of a (small) velocity $\V$ in the aging regime.

\section{Discussion}

We have introduced a spatial asymmetry
into the linear Langevin dynamics for the ferromagnetic spherical model
on the hypercubic lattice.
The asymmetry is measured by an arbitrary velocity vector $\V$.
The resulting dynamics is irreversible.
It therefore breaks detailed balance and its numerous consequences,
including the \fd theorem.
The corresponding nonequilibrium stationary state is however still Gibbsian,
with the weights of configurations being dictated by the ferromagnetic Hamiltonian.

The model remains exactly solvable in any dimension,
allowing thus an analytical evaluation of time-dependent observables.
The main emphasis has been put on two-time quantities,
and especially on the fluctuation-dissipation ratio.
We have performed a systematic investigation of several regimes of interest,
either stationary or transient,
for arbitrary dimensions and in the different phases of the model.
One of the most noticeable outcomes of this study
is the existence of two regimes of violation of the \fd theorem
in the nonequilibrium stationary state:
a regime of weak violation at small $\V$,
where the stationary value of the \fd ratio is finite
but less than unity, and varies continuously with $\V$,
and a regime of strong violation at large $\V$,
where the \fd ratio vanishes asymptotically.
This phenomenon had been uncovered for the first time
in the kinetic Ising chain under asymmetric dynamics~\cite{cg2011}.
The present study suggests that this novel kind of dynamical transition
in nonequilibrium stationary states might be quite general.
We have also characterized the rounding of the above transition
in the transient regime.

As mentioned in the introduction, numerical simulations of the two-dimensional Ising model
with the asymmetric dynamics introduced in~\cite{gb2009}
are currently in progress~\cite{gp}.
Finally, it would be worth pursuing the exploration
of irreversible dynamics and nonequilibrium stationary states
driven by spatial asymmetries
in other directions, such as conserved dynamics,
or with other approaches, such as field-theoretical methods.

\subsection*{Acknowledgments}

It is a pleasure to thank Malte Henkel for interesting discussions
during preliminary stages of this work.

\end{document}